\documentclass[
aps,                    % American Physical Society styling. Default.
pre,
twocolumn,              % Two column formatting.
10pt,                   % Set font size. [preprint] gives [12pt], [twocolumn]
superscriptaddress,     % Associate authors with affiliations via superscript
preprintnumbers,        % Control display of preprint numbers given by
nofootinbib,             % journal style.
showpacs,               % Control display of PACS: line.
showkeys,               % Control display of Keywords: line.
]{revtex4-2}

\usepackage{bigdelim}

\usepackage{graphicx}               % include graphics
\usepackage[caption=false]{subfig}  % subfigures, don't load captions for
\usepackage{ragged2e}               % for the \justifying macro
\DeclareCaptionJustification{justified}{\justifying}
\usepackage[svgnames]{xcolor}       % colors 
\usepackage{xspace}                 % spaces in macros

\usepackage[T1]{fontenc}                 % allow copy/paste of unicode
\usepackage{lmodern}                     % nicer looking font
\usepackage[scaled=.8]{beramono}         % more nicer fonts
\usepackage{microtype}                   % friendlier word spacing

\usepackage{dcolumn}                    % organize columns by decimal point
\usepackage{booktabs}                   % nicer table horizontal rules
\usepackage{longtable}                  % allow tables to flow over page breaks

\usepackage{old-arrows}                 % make overarrows look a little less
\usepackage{url}                        % format URLs
\usepackage[super]{nth}                 % 1st, 2nd, etc using superscripts
\usepackage{lipsum}                     % create lorum ipsum
\usepackage{siunitx}                    % format units including bits
\usepackage{algorithm}
\usepackage{algpseudocode}
\usepackage{csquotes}                   % format quotes

\usepackage{amsmath}        % AMS mathematical typesetting
\usepackage{amsfonts}       % AMS mathematical fonts
\usepackage{amssymb}        % AMS mathematical symbols 
\usepackage{amsthm}         % AMS theorems

\theoremstyle{plain}            
\theoremstyle{plain}            
\theoremstyle{plain}            
\theoremstyle{plain}            
\theoremstyle{plain}            
\theoremstyle{remark}           
\theoremstyle{definition}       \newtheorem{definition}{Definition}

\usepackage{overarrows}                     % building custom arrows
\usepackage{mathtools}                      % mathclap, etc
\usepackage{mathrsfs}                       % mathscr 
\usepackage{upgreek}                        % more greek letters
\usepackage{braket}                         % bras and kets for quantum
\usepackage{stmaryrd}                       % some computer science symbols
\SetSymbolFont{stmry}{bold}{U}{stmry}{m}{n} % set how stmry handles bold
\usepackage{nicefrac}                       % pretty fractions
\usepackage{bm}                             % bold in math mode 
\usepackage{bbm}                            % more bold

\usepackage{hyperref}           % support for hypertext
\hypersetup{
    colorlinks   = true,        % Colours links instead of ugly boxes
    linkcolor={blue!50!black},
    citecolor={blue!50!black},
    urlcolor={blue!80!black}
    }

\usepackage[capitalise]{cleveref}       % easily format refs to figures etc 
\newcommand{\eM}     {\mbox{$\epsilon$-machine}\xspace }
\newcommand{\eMs}    {\mbox{$\epsilon$-machines}\xspace }
 
\newcommand{\EMs}    {\mbox{$\epsilon$-Machines}\xspace }

\NewOverArrowCommand{\PastArrow}{start=\leftharpoonup,
                                 end=\relbar,
                                 trim=6,
                                 space after arrow=-.1ex}

\NewOverArrowCommand{\pastArrow}{start={\smallermathstyle\leftharpoonup},
                                 end=\relbar,
                                 trim start = 5,
                                 trim end = 8,
                                 }
                
\NewOverArrowCommand{\FutureArrow}{end=\rightharpoonup, 
                                   trim=7,
                                   space after arrow=-.1ex}
    
\NewOverArrowCommand{\futureArrow}{start={\smallermathstyle\relbar}, 
                                   end=\rightharpoonup,
                                   trim start = 5,
                                   trim end = 8,
                                  }
                                
\NewOverArrowCommand{\BiArrow}{start=\leftharpoonup,
                                 end=\rightharpoonup,
                                 trim=8,
                                 space after arrow=-.1ex}

\NewOverArrowCommand{\biArrow}{start={\smallermathstyle\leftharpoonup},
                                 end=\rightharpoonup,
                                 trim=9
                                 }

\newcommand{\Machine}               { M }
\newcommand{\eMachine}              { \Machine_{\epsilon} }

\newcommand{\EquiFunction}[1]       { \epsilon \left[ #1\right] }
\newcommand{\Process}               { \mathcal{P} }

\newcommand{\MeasSymbol}            { {X} }
\newcommand{\meassymbol}            { {x} }
\newcommand{\biinfinity}            { \biArrow {\meassymbol} }
\DeclareRobustCommand{\Past}        { \PastArrow{\MeasSymbol} }
\newcommand{\PastSmashed}           { \smash{\Past} }

\newcommand{\past}                  { \pastArrow{\meassymbol} }
\newcommand{\pastprime}             { \past^{\prime} }

\DeclareRobustCommand{\Future}      { \FutureArrow {\MeasSymbol} }
\newcommand{\FutureSmashed}         { \smash{\Future} }

\newcommand{\future}                { \futureArrow {\meassymbol} }
\newcommand{\futureprime}           { \future^{\prime} }

\newcommand{\AllPasts}              { \PastArrow{\bm{\MeasSymbol}} }

\newcommand{\CausalState}           { \mathcal{S} }

\newcommand{\CausalStateSet}        { \bm{\CausalState} }

\newcommand{\causalstate}           { s }

\renewcommand{\H}[1]                { \operatorname{H} \left[ #1 \right] }
\newcommand{\Hcond}[2]              { \operatorname{H} \left[ #1 \mid #2 \right] }  
\newcommand{\I}[2]                  { \operatorname{I} \left[ #1 ; #2 \right] }
\newcommand{\Icond}[3]              { \operatorname{I} \left[ #1 ; #2 \mid #3 \right] } 
\newcommand{\ExcessEntropy}         { \bm{E} }

\newcommand{\Crypticity}            { \chi }

\newcommand{\Cmu}                   { C_{\mu} }

\newcommand{\hmu}                   { h_{\mu} }

\newcommand{\rmu}                   { r_{\mu} }
\newcommand{\qmu}                   { q_{\mu} }
\newcommand{\bmu}                   { b_{\mu} }
\newcommand{\sigmu}                 { \sigma_{\mu} }
\newcommand{\bmuforward}            { \bmu^+ }
\newcommand{\bmureverse}            { \bmu^- }
\newcommand{\integers}              { \mathbb{Z} }
\newcommand{\nats}                  { \mathbb{N} }

\newcommand{\sigmaAlgebra}          { \Sigma }

\newcommand{\powerSet}              { \mathcal{P} }

\newcommand{\measure}               { \mu }

\newcommand{\forward}                   { + }
\newcommand{\reverse}                   { - }
\newcommand{\forwardreverse}            { \pm }

\newcommand{\ForwardCausalState}        { \CausalState^{\forward} }

\newcommand{\ForwardCausalStateSet}     { \CausalStateSet^{\forward} }
\newcommand{\forwardcausalstate}        { \causalstate^{\forward} }

\newcommand{\ReverseCausalState}        { \CausalState^{\reverse} }

\newcommand{\ReverseCausalStateSet}     { \CausalStateSet^{\reverse} }
\newcommand{\reversecausalstate}        { \causalstate^{\reverse} }

\newcommand{\BiCausalState}             { \CausalState^{\forwardreverse} }

\newcommand{\BiCausalStateSet}          { \CausalStateSet^{\forwardreverse} }
\newcommand{\bicausalstate}             { \causalstate^{\forwardreverse} }

\newcommand{\ForwardEM}                 { \eMachine^\forward }

\newcommand{\ReverseEM}                 { \eMachine^\reverse }

\newcommand{\BiEM}                      { \eMachine^\forwardreverse }
\newcommand{\ForwardCrypticity}         { \Crypticity^{\forward} }
\newcommand{\ReverseCrypticity}         { \Crypticity^{\reverse} }

\newcommand{\MixedState}                { \mathcal{H} }

\newcommand{\MixedStateSet}             { \bm{\MixedState} }
\newcommand{\mixedstate}                { \eta }

\newcommand{\RVSet}             { \mathfrak{X} }

\newcommand{\atom}              { \alpha }
\newcommand{\PRVSet}            { \powerSet(\RVSet) }

\newcommand{\shiftOperator}     { \sigma }
\newcommand{\alphabet}          { \mathcal{\MeasSymbol} }

\newcommand{\Present}           { \MeasSymbol_0 }

\newcommand{\MSym}              { \MeasSymbol}

\newcommand{\Prob}              { \Pr} % use standard command
\newcommand{\MS}[2]             { \ensuremath{\MeasSymbol_{#1:#2}}\xspace}

\newcommand{\MeasSymbolVar}         {Y}
\newcommand{\meassymbolvar}         {y}
\newcommand{\MeasSymbolVarAlphabet} {\mathcal{\MeasSymbolVar}}
\newcommand{\MeasSymbolVarTwo}      {Z}

\newcommand{\Asubset}{A}

\newcommand{\typeOne}{transient\xspace }
\newcommand{\typeTwo}{persistent\xspace }

\newcommand{\sF}{\mathcal{F}}
\newcommand{\GenSet}{F}
\newcommand{\Imeasure}{\mu^*}

\newcommand{\word}{w}
\newcommand{\flipped}[1]{\widetilde{#1}}
\newcommand{\SwitchingM}{S}
\newcommand{\ForwardS}{\SwitchingM^+}
\newcommand{\ReverseS}{\SwitchingM^-}
\newcommand{\ForwardA}{A^+}
\newcommand{\ReverseA}{A^-}
\newcommand{\numFStates}{N}
\newcommand{\numRStates}{M}

\usepackage{tabularx}

\makeatletter
\newcommand{\multiline}[1]{%
  \begin{tabularx}{\dimexpr.9\columnwidth-\ALG@thistlm}[t]{@{}X@{}}
    #1
  \end{tabularx}
}
\makeatother

\begin{document}

\title{Taxonomy of Prediction}

\author{Alexandra Jurgens}
\email{alexandra.jurgens@inria.fr}
\homepage{http://csc.ucdavis.edu/~ajurgens/}
\affiliation{GEOSTAT Team,
             INRIA -- Bordeaux Sud Ouest \\
             33405 Talence Cedex, France}

\author{James P. Crutchfield}
\email{chaos@ucdavis.edu}
\homepage{http://csc.ucdavis.edu/~chaos/}
\affiliation{Complexity Sciences Center and Physics Department \\ 
             University of California at Davis,
             One Shields Avenue, Davis, CA 95616}

%%%%%%%%%%%%%%%%%%%%%%%%%%%%%%%%%%%%%%%%%%%%%%%%%%%%%%%%%%%%%%%%%%%%%%%%%%%%%%%

\begin{abstract} 
A prediction makes a claim about a system's future given knowledge of its past.
A retrodiction makes a claim about its past given knowledge of its future. The
bidirectional machine is an ambidextrous hidden Markov chain that does both
optimally by making explicit in its state structure all statistical correlations
in a stochastic process. We introduce an informational taxonomy to profile these
correlations via a suite of multivariate information measures. While prior
results laid out the different kinds of information contained in isolated
measurement of a bit, the associated informations were challenging to calculate
explicitly. Overcoming this via bidirectional machine states, we expand that
analysis to prediction and retrodiction. The result highlights fourteen new
interpretable and calculable measures that characterize a process' informational
structure. In addition, we introduce a labeling and indexing scheme that
systematizes information-theoretic analyses of complex multivariate systems.
Operationalizing this, we provide algorithms to directly calculate all of these
quantities in closed form for finitely-modeled processes.
\end{abstract}

%%%%%%%%%%%%%%%%%%%%%%%%%%%%%%%%%%%%%%%%%%%%%%%%%%%%%%%%%%%%%%%%%%%%%%%%%%%%%%%

\date{\today}

\preprint{arxiv.org:2504.11371 [cond-mat.stat-mech]}
\keywords{entropy rate, multivariate mutual information, information
diagram, information atoms}

\maketitle
\tableofcontents

%%%%%%%%%%%%%%%%%%%%%%%%%%%%%%%%%%%%%%%%%%%%%%%%%%%%%%%%%%%%%%%%%%%%%%%%%%%%%%%

\section{Introduction}
\label{sec:introduction}

How much information can be learned from a single measurement? Shannon
information theory tells us that, on average, information learned by observing a
single realization of a random variable is equivalent to the reduction in our
uncertainty over the outcome \cite{Shan48a,Cove06a,Yeun08a}. This means that
more information is learned from a fair coin flip than from the outcome of a
biased one, and indeed the amount of information is proportional to the bias,
going to zero when heads or tails becomes certain.

What about a sequence of measurements? If the coin flip is one in a sequence of
identical coin flips---an independent identically-distributed (IID)
sequence---the answer is simple: each successive measurement gives the same
amount of information. However, here we are not interested in analyzing IID
sequences, but rather those that have structure in the form of correlations
across time.

Measurement of this kind of statistical dependence has a long history, going
back at least to the 1700s with Jacob Bernoulli \cite{Bern13a} and the 1800s
with Simeon Poisson \cite{Pois37a} and Pafnuty Chebyshev \cite{Cheb67a}. Its
more modern form, though, was initially developed by Andrei Andreevich Markov
\cite{Mark08a} at the turn of the $20^\mathrm{th}$ century. These culminated in
the weak Law of Large Numbers, the Central Limit Theorem, and Markov
chains---transition probabilities, irreducibility, and stationarity---to
mention only a few of the concepts we use today.

In the 1940s, Shannon introduced the \emph{entropy rate} of stationary,
discrete symbol and discrete time processes, which quantifies how much new
information we learn upon successive observations, given knowledge
of their infinite past \cite{Shan48a}. Or, to change the question around, how
\emph{predictable} the new measurement is given knowledge of the history. This
was the first Shannon information measure developed to describe the presence of
temporal correlational structure in a stochastic process in terms of its
relationship to a single measurement. (Shannon's first application, highly
relevant to the modern day, was to predicting natural language \cite{Shan51a}.)

Today, an extensive suite of information rates has been developed to identify
the kinds of information in a single measurement of a stochastic process---the
``anatomy of a bit'' \cite{Jame11a}. This anatomy includes five distinct
information measures that describe the information contained in a single bit in
terms of its correlational relationship to a process' past and future. 

The following expands this anatomy by analyzing not just an isolated bit's
relationship to the past and future, but also to an optimally predictive model
of the stochastic process; i.e., to one whose error rate is bounded below by the
process' Shannon entropy rate \cite{Cove06a}. When also constrained to be
minimal, the optimally predictive model is unique and is a hidden Markov chain
(HMC) called the \eM \cite{Shal98a}. The \eM necessarily captures in its state
structure all information in the process required for optimal prediction---which
is to say, the long-range historical correlations that impact the future. Thus,
in building a full information taxonomy of prediction we analyze the information
present in not only the bits that the model predicts but also the states of the
model itself.

Notably, expanding the analysis into a complete taxonomy also requires the
equivalent but complementary task of retrodiction---making claims about the
past given the present. Although it is well known that the Shannon entropy rate
is time-symmetric for stationary processes, the tasks of optimal prediction and
optimal retrodiction are not. Prediction and retrodiction generically require
different modeling architectures, even for relatively simple discrete
processes. Specifically, to characterize the informational structure of
prediction, one needs to consider not only the architecture of the predictive
``forward time'' \eM but also the architecture of the retrodictive ``reverse
time'' \eM. These architectures capture correlations in the process that impact
the present but are not accessible through isolated measurements. 

To this end, the following invokes the \emph{bidirectional machine}, an
ambidextrous hidden Markov chain capable of simultaneous optimal prediction and
retrodiction \cite{Crut08a}. We show that knowledge of the bidirectional machine
allows one to fully characterize a prediction---which we take to be the
observation \emph{and} all inaccessible but relevant information in the
process---using a taxonomy of fourteen information quantities. Furthermore, and
importantly, we show that these are exactly calculable in closed form and do not
need to be approximated as the limits of information rates, as previously.

Given that this setting involves highly multivariate information ($n$-way
correlations across arbitrary times), we first review information theory. We
then introduce a systematic method for generating the set of ``irreducible''
information atoms for an arbitrary set of random variables. We apply this to a
single prediction (or retrodiction) of the bidirectional machine, generating
fourteen informational atoms that describe the full informational structure of
the model's average prediction (or retrodiction). We then relate these atoms to
previously-defined information measures, resulting in an informational taxonomy
of prediction. Finally, we give several worked examples for binary stochastic
processes of increasing complexity, along with the algorithms needed.

\section{Information Theory}
\label{sec:Background}

To study and characterize processes and their associated models we make use of
\emph{Shannon's information theory} \cite{Shan48a,Cove06a,Yeun08a}, a
widely-used foundational framework that provides tools to describe how
stochastic processes generate, store, and transmit information. First, though,
we deviate some from our main technical development to briefly recall several
basic concepts it requires. The reader familiar with information theory may
comfortably skip this section, although the notation given in
\cref{sec:infoAtomConstruction} for finding sets of information atoms of
arbitrary random variables will be useful later on. 

\subsection{Measures}
\label{sec:InformationTheory}

Let $\MeasSymbol$ be a discrete-valued random variable defined on a
\emph{probability space} $\left( \alphabet, \sigmaAlgebra, \measure \right)$
\cite{Kall01a,Kurk03a}. We call $\alphabet$ the \emph{event space} or
\emph{measurement alphabet} of $\MeasSymbol$ and take it to be a finite set.
The probability of random variable $X$ taking value $x$ is determined by the
\emph{measure} $\mu$: $\Pr \left( \MeasSymbol = \meassymbol \right) = \measure
\left( \left\{ \meassymbol \right\} \in \alphabet \right)$. That is, we denote
instances of random variables by capital Latin letters and specific
realizations by lower case.

The most basic quantity in information theory is the \emph{Shannon
entropy}---the average amount of information learned upon a single measurement
of a random variable. (It is, modulo sign, also the amount of uncertainty one
faces when predicting the outcome of the measurement.) The Shannon entropy
$\H{\MeasSymbol}$ of the random variable $\MeasSymbol$ is defined:
\begin{align}
    \H{\MeasSymbol} =
    - \sum_{\meassymbol \in \alphabet} 
    \Pr(\MeasSymbol = \meassymbol)
    \log_2 \Pr(\MeasSymbol = \meassymbol)
    ~.
\label{eq:ShannonEntropy}
\end{align}

We can also characterize the relationship between a pair of jointly-distributed
random variables, say, $\MeasSymbol$ and $\MeasSymbolVar$. The \emph{joint
entropy} $\H{\MeasSymbol, \MeasSymbolVar}$ is of the same functional form as
\cref{eq:ShannonEntropy}, applied to the joint distribution $\Pr \left(
\MeasSymbol, \MeasSymbolVar \right)$. This can, in principle, be
straightforwardly extended to a set of $N$ variables $\RVSet = \left\{
\MeasSymbol_i \mid i \in \left( 1, \dots, N \right) \right\}$.

The \emph{conditional entropy} $\Hcond{\MeasSymbol}{\MeasSymbolVar}$ gives the
additional information learned from observation of one random variable
$\MeasSymbol$ given knowledge of another random variable $\MeasSymbolVar$. The
conditional entropy is given by:
\begin{align}
    \Hcond{\MeasSymbol}{\MeasSymbolVar} = 
    \H{\MeasSymbol, \MeasSymbolVar} - \H{\MeasSymbolVar}  ~. 
    \label{eq:ConditionaLEntropy}
\end{align}

% The chain rule relating the joint Shannon entropy of $N$ variables to a sum of
% conditional entropies is given by:
% \begin{align*}
%     \H{\MeasSymbol_1, \MeasSymbol_2, \dots , \MeasSymbol_N} = 
%     \sum_{i}^N \Hcond{\MeasSymbol_i}{\MeasSymbol_1, \dots, \MeasSymbol_{i-1}} ~. 
% \end{align*}

The fundamental measure of information shared between random variables is the
\emph{mutual information}:
\begin{align}
    \I{\MeasSymbol}{\MeasSymbolVar} = 
    \sum_{ \meassymbolvar \in \MeasSymbolVarAlphabet } 
    \sum_{\meassymbol \in \alphabet} & \Pr \left( \MeasSymbol = \meassymbol, 
    \MeasSymbolVar = \meassymbolvar \right) \times \nonumber \\ 
    & \log_2  \left( \frac{\Pr \left( \MeasSymbol = \meassymbol, 
    \MeasSymbolVar = \meassymbolvar \right) }{\Pr 
    \left( \MeasSymbol = \meassymbol \right) 
    \Pr \left( \MeasSymbolVar = \meassymbolvar \right)} \right) ~. 
\end{align}
The probabilities of both variables are taken over the joint probability
distribution, while the single probabilities are taken according to the
marginals. The mutual information can also be written in terms of Shannon
entropies and conditional entropies:
\begin{align}
    \I{\MeasSymbol}{\MeasSymbolVar} & = 
    \H{\MeasSymbol, \MeasSymbolVar} - \Hcond{\MeasSymbol}{\MeasSymbolVar} - 
    \Hcond{\MeasSymbolVar}{\MeasSymbol} \nonumber \\ 
    & = \H{\MeasSymbol} + \H{\MeasSymbolVar} - \H{\MeasSymbol, \MeasSymbolVar} ~. 
    \label{eq:MutualInformation}
\end{align}

Direct inspection shows that the mutual information between two variables is
symmetric. The mutual information between vanishes if and only if  $\MeasSymbol$
and $\MeasSymbolVar$ are statistically independent.

As with entropy, we may condition the mutual information on another random
variable $\MeasSymbolVarTwo$, giving the \emph{conditional mutual information}:
\begin{align}
    \Icond{\MeasSymbol}{\MeasSymbolVar}{\MeasSymbolVarTwo} =
    \Hcond{\MeasSymbol}{\MeasSymbolVarTwo} + 
    \Hcond{\MeasSymbolVar}{\MeasSymbolVarTwo} - 
    \Hcond{\MeasSymbol, \MeasSymbolVar}{\MeasSymbolVarTwo} ~.
    \label{eq:condMI}
\end{align}
The conditional mutual information is the amount of information shared by
$\MeasSymbol$ and $\MeasSymbolVar$, given we know the third $\MeasSymbolVarTwo$.

Similar to the joint entropy, the mutual information between all three
variables---also known as the \emph{interaction information} or the
\emph{multivariate mutual information}---is given by the difference between
mutual information and conditional mutual information:
\begin{align}
    \operatorname{I} \left[ \MeasSymbol ; \MeasSymbolVar ; 
    \MeasSymbolVarTwo \right] = 
    \I{\MeasSymbol}{\MeasSymbolVar} - 
    \Icond{\MeasSymbol}{\MeasSymbolVar}{\MeasSymbolVarTwo} ~. 
    \label{eq:interactionInfo}
\end{align}

There are two cases worth pointing out here. Two variables $\MeasSymbol$ and
$\MeasSymbolVar$ can have positive mutual information but be conditionally
independent in the presence of $\MeasSymbolVarTwo$, in which case the
interaction information is positive. It is also possible, though, for two
independent variables to become correlated in the presence of
$\MeasSymbolVarTwo$, making the conditional mutual information positive and the
interaction information negative. In other words, conditioning on a third
variable $\MeasSymbolVarTwo$ can either increase or decrease mutual information
and $\MeasSymbol$ and $\MeasSymbolVar$ variables can appear more or less
dependent given additional data \cite{Cove06a}. That is, there can be
\emph{conditional independence} or \emph{conditional dependence} between a pair
of random variables. Note that the interaction information is symmetric, so this
intuition holds regardless of the conditioning variable selected. 

\subsection{Diagrams}
\label{sec:InformationDiagrams}

\begin{figure}[t] \centering
    \includegraphics[width=\columnwidth]{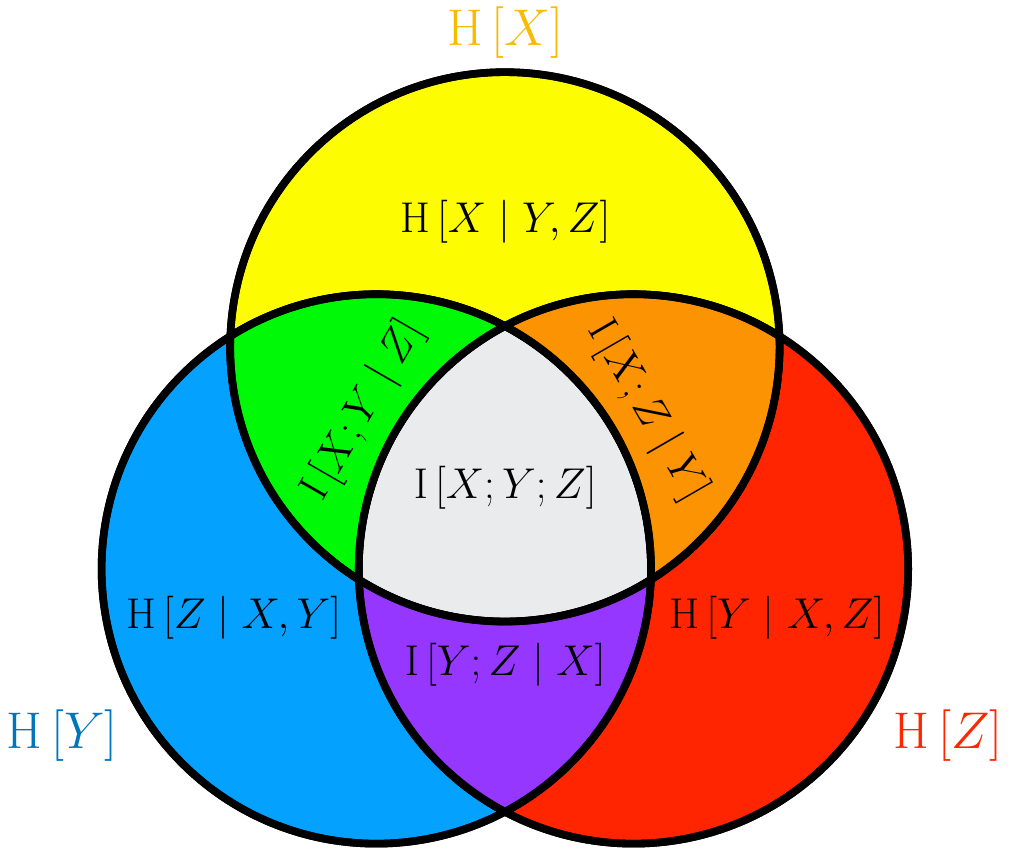}
    \caption{ Information diagram with three random variables, $\MeasSymbol$,
    $\MeasSymbolVar$, and $\MeasSymbolVarTwo$. }
    \label{fig:iDiagramThreeVariable} 
  \end{figure}

We will now make the relationship between information quantities defined in the
last section and the algebra of events clear \cite{Yeun08a}. First, consider only
two random variables $\MeasSymbol$ and $\MeasSymbolVar$. The set of the
associated event spaces $\RVSet = \left\{ \alphabet, \MeasSymbolVarAlphabet
\right\}$ induces an algebra $\sF$ over $\RVSet$ closed under complements,
unions, and intersections. $\sF$ is generated by the partition:
\begin{align*}
    \GenSet = \left\{ \alphabet \setminus
    \MeasSymbolVarAlphabet, \MeasSymbolVarAlphabet \setminus \alphabet, \alphabet
    \bigcap \MeasSymbolVarAlphabet , \Omega \setminus \left( \alphabet \bigcup
    \MeasSymbolVarAlphabet \right) \right\} ~.  
\end{align*}
The backslash is set subtraction and $\Omega$ is the universal set.
 
Note that these elements correspond to the unique areas of an Euler diagram of
two overlapping but nonidentical and nonempty sets. The algebra $\sF$ over
$\RVSet$ is generated by unions over $\GenSet$ and so has $2^{|\GenSet]} = 2^4 =
16$ elements. We will discuss the case of arbitrarily many variables in the next
section, but in general for $N$ variables $| \GenSet | = 2^N$ and $| \sF | =
2^{2^N} $. 

Now, specify a real-valued measure $\Imeasure$ for each element in $F$ such
that: 
\begin{enumerate}
      \setlength{\topsep}{-4pt}
      \setlength{\itemsep}{-4pt}
      \setlength{\parsep}{-4pt}
    \item $\Imeasure(\alphabet \setminus \MeasSymbolVarAlphabet) = 
          \Hcond{\MeasSymbol}{\MeasSymbolVar}$,
    \item $\Imeasure(\MeasSymbolVarAlphabet \setminus \alphabet) = 
          \Hcond{\MeasSymbolVar}{\MeasSymbol}$,
    \item $\Imeasure(\alphabet \bigcap \MeasSymbolVarAlphabet) =
          \I{\MeasSymbol}{\MeasSymbolVar}$,
    \item $\Imeasure\left( \Omega \setminus \left( \alphabet \bigcup
          \MeasSymbolVarAlphabet \right) \right) = 
          \Imeasure(\emptyset) = 0$.
\end{enumerate}

It has been shown that $\Imeasure$ exists and corresponds uniquely to the joint
probability measure on $\MeasSymbol$ and $\MeasSymbolVar$ \cite{Ting62a}. In
other words, information can be reframed as an additive set function, revealing
that there is no semantic difference between ``types'' of information---entropy,
mutual information, and so on---but rather a single underlying quantity being
referenced. We call the elements of $\sF$ \emph{information atoms}. The elements
of $\GenSet$ cannot be decomposed into a sum of other information atoms and
are so called the \emph{irreducible} atoms. They circumscribe the range of
possible correlations detectable by Shannon entropies between random variables
in a set. (See Ref. \cite{Jame16a} for examples of multivariate dependence that
are not.)

The correspondence between information and the event algebra allows us to
represent information quantities via an \emph{information diagram}---an Euler
diagram representing the informational relationships between variables. The
entropies of some number of random variables---$\H{\MeasSymbol}$,
$\H{\MeasSymbolVar}$, $\H{\MeasSymbolVarTwo}$, and so on---are represented by
the area contained in their respective circle. A three-variable example is
shown in \cref{fig:iDiagramThreeVariable}. When two variables are independent,
their respective circles do not overlap. Conditioning corresponds to area
subtraction, and shared information to area intersection. 

Information diagrams are useful graphical tools but note that they may be
misleading---$\Imeasure$ is a signed measure, but all nonzero atoms are
visually portrayed by the i-diagram as having positive area. It is also possible
for the informational quantity depicted by an i-diagram to diverge---for
instance, the joint entropy of infinitely many random variables---such as in the
stochastic processes we will encounter. Furthermore, it is difficult to
practically use i-diagrams beyond five or six random variables (unless those
random variables have helpful relational structure that limits the size of
$\sF$). Despite these limitations, i-diagrams remain the tool of choice for
visualizing information-theoretic structure in collections of random variables. 

\subsection{Collections of Variables}
\label{sec:infoAtomConstruction}

\begin{table*}[]
    \def\arraystretch{1.5}
    \begin{tabular}{ccc
        @{\hskip 0.5in}rl@{\hskip 0.5in}
        D{|}{\mid}{-1}}
    \toprule
    \multicolumn{3}{c}{Label Type $\hspace{2.5em}$} & 
    \multicolumn{2}{c}{Partition $\hspace{2.5em}$}  & 
    \multicolumn{1}{c}{Information Atom}  \\
    \toprule
    Decimal & 
    \multicolumn{1}{c}{ Lexicographic } &
    Indicial  & 
    Joint Dist. & 
    Conditioned & \\
    % \midrule
    $i$ & 
    $\hspace{.5em}\MeasSymbol \hspace{.3em}\MeasSymbolVar 
    \hspace{.3em}\MeasSymbolVarTwo\hspace{.5em}$ & 
    $k$ & 
    \multicolumn{1}{r}{$ \Asubset_i $} &
    \multicolumn{1}{l}{$ \overline{\Asubset_i} $} & 
    \multicolumn{1}{c}{$ \atom_i$} \\
    \midrule
    1 & 1 $\hspace{0.2em}$ 0 $\hspace{0.2em}$ 0 & 0 &
    $ \left\{ \MeasSymbol  \right\} $ &
    $ \left\{ \MeasSymbolVar, \MeasSymbolVarTwo  \right\} $  & 
    \operatorname{H} \left[\MeasSymbol \right. | 
    \left. \MeasSymbolVar, \MeasSymbolVarTwo \right] 
    \\ 
    2 & 0 $\hspace{0.2em}$ 1 $\hspace{0.2em}$ 0 & 1 &
    $ \left\{ \MeasSymbolVar  \right\} $ &
    $ \left\{ \MeasSymbol, \MeasSymbolVarTwo  \right\} $  & 
    \operatorname{H} \left[\MeasSymbolVar \right. | 
    \left. \MeasSymbol, \MeasSymbolVarTwo \right] 
    \\ 
    3 & 1 $\hspace{0.2em}$ 1 $\hspace{0.2em}$ 0 & 01 &
    $ \left\{ \MeasSymbol, \MeasSymbolVar  \right\} $ &
    $ \left\{ \MeasSymbolVarTwo  \right\} $  & 
    \operatorname{I} \left[\MeasSymbol; \MeasSymbolVar \right. | 
    \left. \MeasSymbolVarTwo \right] 
    \\ 
    4 & 0 $\hspace{0.2em}$ 0 $\hspace{0.2em}$ 1 & 2 &
    $ \left\{ \MeasSymbolVarTwo  \right\} $ &
    $ \left\{ \MeasSymbol, \MeasSymbolVar  \right\} $  & 
    \operatorname{H} \left[\MeasSymbolVarTwo \right. | 
    \left. \MeasSymbol, \MeasSymbolVar \right] 
    \\ 
    5 & 1 $\hspace{0.2em}$ 0 $\hspace{0.2em}$ 1 & 02 &
    $ \left\{ \MeasSymbol, \MeasSymbolVarTwo  \right\} $ &
    $ \left\{ \MeasSymbolVar  \right\} $  & 
    \operatorname{I} \left[\MeasSymbol; \MeasSymbolVarTwo \right. | 
    \left. \MeasSymbolVar \right] 
    \\ 
    6 & 0 $\hspace{0.2em}$ 1 $\hspace{0.2em}$ 1 & 12 &
    $ \left\{ \MeasSymbolVar, \MeasSymbolVarTwo  \right\} $ &
    $ \left\{ \MeasSymbol  \right\} $  & 
    \operatorname{I} \left[\MeasSymbolVar; \MeasSymbolVarTwo \right. | 
    \left. \MeasSymbol \right] 
    \\ 
    7 & 1 $\hspace{0.2em}$ 1 $\hspace{0.2em}$ 1 & 012 &
    $ \left\{ \MeasSymbol, \MeasSymbolVar, \MeasSymbolVarTwo  \right\} $ &
    $ \emptyset $  & 
    \operatorname{I} \left[\MeasSymbol; 
    \MeasSymbolVar; \MeasSymbolVarTwo \right] 
    \\ 
    \bottomrule
    \end{tabular}
\caption{The irreducible information atoms for a set of three random variables
	$\RVSet = \left\{ \MeasSymbol, \MeasSymbolVar, \MeasSymbolVarTwo \right\}$.
	Compare the list of $\alpha_i$ to the areas of the information diagram
	depicted in \cref{fig:iDiagramThreeVariable}.
	}
\label{tab:InfoAtomsThreeVariable}
\end{table*}

Now, we will show how to find $\GenSet$, $\sF$, and $\Imeasure$ for an
arbitrary collection of random variables $\RVSet = \left\{ \MeasSymbol_0,
\MeasSymbol_1, \dots, \MeasSymbol_k, \dots, \MeasSymbol_{N-1} \right\}$. To be
explicit when taking functions of sets, we borrow the iterable unpacking
notation common in modern programming languages. So, we write:
\begin{align*}
    f(* \Asubset) = 
    f \left(\MeasSymbol_0, \MeasSymbol_1, \dots, 
    \MeasSymbol_k, \dots, \MeasSymbol_{N-1} \right)
\end{align*}
where $\Asubset = \left\{\MeasSymbol_0, \MeasSymbol_1, \dots, \MeasSymbol_k,
\dots, \MeasSymbol_{N-1} \right\}$. We also abuse notation and take all
power sets to exclude the empty set by default; i.e., $\powerSet (\RVSet) =
\powerSet (\RVSet) \setminus \emptyset$. With this notation we concisely write
down the interaction information for arbitrary variables as:
\begin{align}
    \operatorname{I} \left[ * \RVSet \right] 
    & = \sum_{\Asubset \in \PRVSet} \left(-1\right)^{\vert \Asubset \vert - 1} 
    \H{*\Asubset} ~. 
    \label{eq:multivariateMI}
\end{align}
(Compare to \cref{eq:interactionInfo}, \cref{eq:condMI}, and
\cref{eq:MutualInformation}.)

The challenge is to construct the set of irreducible information atoms for a
finite random variable set $\RVSet$ of size $N$. This set consists of,
maximally, $N$ conditional informations, one multivariate mutual information,
$2^N - 2 - N$ conditional mutual informations, and the empty set.

First, there is the arbitrary conditional entropy, which breaks down into two
entropies:
\begin{align}
    \Hcond{*\Asubset}{* \left( \RVSet \setminus \Asubset \right)} =
    \H{* \RVSet} - \H{* \left( \RVSet \setminus \Asubset \right)}
	~,
    \label{eq:NcondH}
\end{align}
where $\Asubset \in \PRVSet$---the power set. 
Then, the arbitrary conditional mutual information is: 
\begin{align}
    \operatorname{I} & 
    \left[ *\Asubset \mid * \left( \RVSet \setminus \Asubset \right) \right] 
     = \nonumber \\
    & \sum_{a \in \powerSet(\Asubset)} \left( -1 \right)^{\vert a \vert + 1}
    \biggl( \H{* \left( a \cup \RVSet \setminus \Asubset \right) } - 
    \H{ * \left( \RVSet \setminus \Asubset \right) }
    \biggr)
	~.
    \label{eq:NcondMI}
\end{align} 

Notice that when $\vert \Asubset \vert = 1$, \cref{eq:NcondMI} reduces to
\cref{eq:NcondH} and when $\Asubset = \RVSet$ it reduces to \cref{eq:multivariateMI}.
So, we only need to apply \cref{eq:NcondMI} to each subset $\Asubset \in
\powerSet(\RVSet)$ find every irreducible information atom---this is equivalent
to finding $\Imeasure(\GenSet)$.

\subsection{Irreducible Information Atoms}
\label{sec:infoatomLabeling}

Working with information atoms for arbitrarily many variables very quickly
becomes unwieldy due to the exponential growth of the number of atoms.
Fortunately, there is a natural ordering for the set of irreducible information
atoms. The atoms are labeled by indexing the power set of $\RVSet$ with an
isomorphism to the binary representation of numbers from $1$ to $2^N - 1$. We
simply indicate the presence of the $\MeasSymbol_k$ variable in a subset by the
$k$th digit of the binary sequence---$1$ if the variable is in the joint
distribution and $0$ if it is being conditioned on. Recall we exclude the empty
set by default. 

Notice that this ordering of binary digits is \emph{reversed} compared to the
typical representation---compare the Lexicographic column in
\cref{tab:InfoAtomsThreeVariable} to the Decimal column. This is due to our
primarily working with time-indexed variables and our choosing (rather
arbitrarily) to imagine time flowing from left-to-right. Ordering the
lexicographic labels from left to right allows easily identifying the semantic
meaning of binary strings at a glance.

Given $i \in [1, \dots, 2^N -1]$, let $\Asubset_i$ be the $i$th set in
$\PRVSet$. The associated irreducible information atom is: 
\begin{align}
    \atom_i = 
    \operatorname{I} \left[ * \Asubset_i \mid * (\RVSet \setminus \Asubset_i) \right] 
	~,
    \label{eq:generalInfoFunction}
\end{align}

So, the set of irreducible information atoms for $\RVSet$ is given by: 
\begin{align}
  \GenSet_\RVSet = 
  \left\{  \atom_i \mid i \in [1, \dots, 2^N -1] \right\} ~. 
  \label{eq:infoAtomOperator}
\end{align}

The explicit listing of $\GenSet$ is given for the $N = 3$ case by
\cref{tab:InfoAtomsThreeVariable}, which also gives the \emph{indicial} label of
each information atom. This is simply the indices $k$ of the random variables
present in the joint distribution. This label is shorter than the lexicographic
and often easier to identify at a glance. It is also useful when the index of
the random variable carries relational meaning, as it will in our specific use
case. The associated i-diagram is depicted in \cref{fig:iDiagramThreeVariable}. 

This completes our review of basic information theory---a toolset
to initiate a full information-theoretic analysis of any set of random
variables if we so chose. In principle, one only need construct $\GenSet$ as
detailed above and then generate the full set of information atoms $\sF$ through
unions. In practice---even assuming one already has access to the full
joint probability distribution over all variables, a nontrivial assumption to
say the least---the growth rate of these sets and the difficulty of mechanistic
interpretation once one begins to consider more than three variables has
historically stymied these approaches. Moreover, the literature has long debated the
semantic meaning of various information atoms---the negativity of interaction
information, to pick one example, has been a hotly-debated topic \cite{Beer14a}.

We sidestep these concerns to a degree by narrowing our focus from a totally
arbitrary collection of random variables to a collection of random variables
that are measurements of a stochastic process over time. This introduces a
significant amount of structure into the informational relationships between
the variables, as we will show in the next section. 

\section{Information in Stochastic Processes}
\label{sec:InfoTheoryofSP}

As noted at the end of \cref{sec:infoAtomConstruction}, we are interested here
not in truly arbitrary collections of random variables but rather stochastic
processes, which are understood as a sequences of random variables related to
each other through time by a particular dynamic. Specifically, we investigate
the relationship between the informational quantities of random variable blocks
belonging to the process and to the process dynamic. 

\subsection{Discrete Discrete Processes}
\label{sec:Processes}

\begin{figure}
\includegraphics[width=\columnwidth]{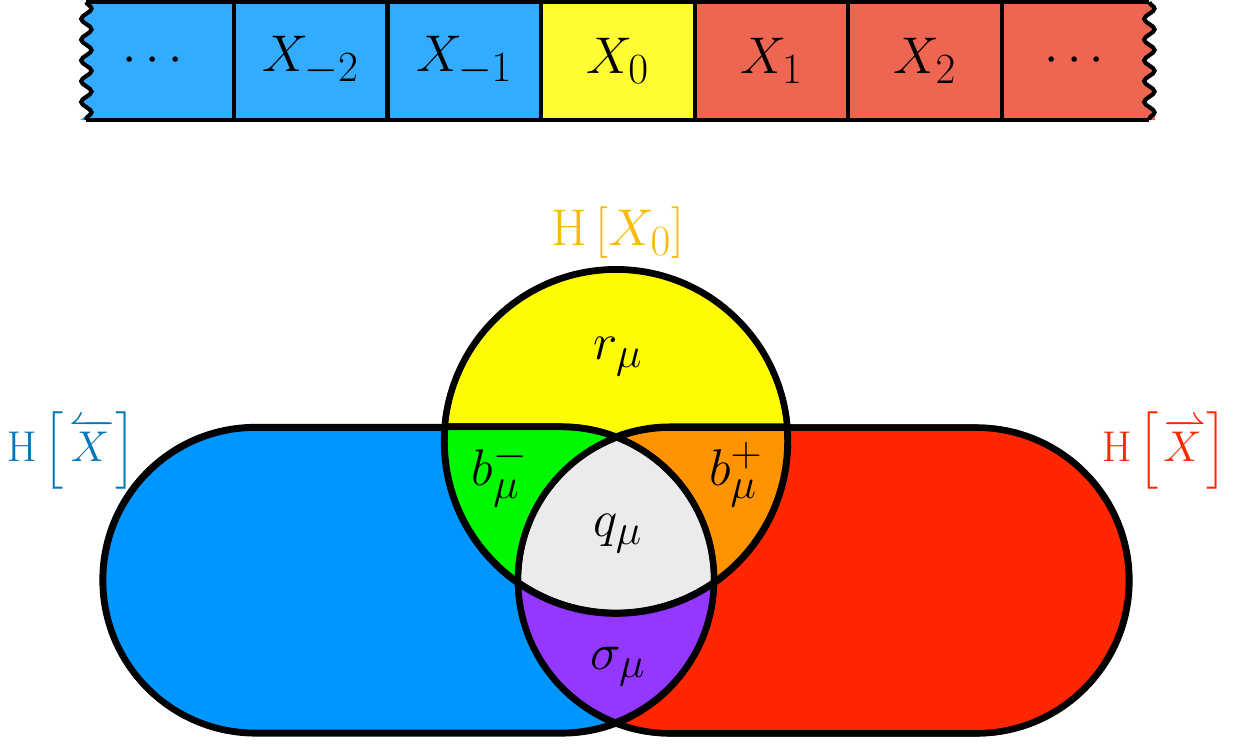}
\caption{(Above) A tape representing a series of measurements of a discrete-time
    stochastic process. (Below) information diagram representing the
    informational relationships between the future $\Future$, the present
	$\Present$, and the past $\Past$ measurements of a generic discrete-symbol,
	discrete-time stochastic process. The i-diagram is labeled atoms: ephemeral
	information $\rmu$, binding informations $\bmu$, enigmatic information
	$\qmu$, and elusive information $\sigmu$.
	}
\label{fig:iDiagramProcess}
\end{figure}

We take a \emph{stochastic process} $\Process$ to consist of a
$\integers$-indexed random variable $\MeasSymbol$ defined on the measure space
$\left( \alphabet^{\integers}, \sigmaAlgebra, \measure \right)$. This indexing
is temporal and is done by the use of subscripts. For example, we write
$\MeasSymbol_t = \meassymbol$ to say that $\meassymbol \in \alphabet$ is the
specific value of $\MeasSymbol$ at time $t$.

The dynamic of the stochastic process is given by the \emph{shift operator},
also called the translation operator, which is an operator $\shiftOperator :
\alphabet^{\integers} \to \alphabet^{\integers}$ that maps $t$ to $t+1$:
$\shiftOperator \meassymbol_t = \meassymbol_{t+1}$. It also acts on the measure:
$ (\shiftOperator \measure) (E) = \measure (\shiftOperator^{-1} E ) $ for $E \in
\sigmaAlgebra$. This addition extends the measure space to a dynamical system
$(\alphabet^{\integers}, \sigmaAlgebra, \measure, \shiftOperator)$.

Blocks of the process, called \emph{words}, are denoted by
$\MeasSymbol_{a:b}  = \left\{ \MeasSymbol_t : a < t \leq b; a, b \in \integers
\right\}$ with the left index inclusive and the right exclusive. A word could
also refer to a particular realization of a given length. For instance, one
might write $\MeasSymbol_{0:3} = \MeasSymbol_0 \MeasSymbol_1 \MeasSymbol_2$
or $\meassymbol_{0:3} = \meassymbol_0 \meassymbol_1 \meassymbol_2$. 

To simplify our mathematical development, we restrict to stationary, ergodic
processes: those for which $\Prob(\MS{t}{t+\ell}) = \Prob(\MS{0}{\ell})$ for all
$t \in \mathbb{\MeasSymbolVarTwo}$, $\ell \in \mathbb{\MeasSymbolVarTwo}^+$, and
for which individual realizations obey all of those statistics.

We refer to the observation at $t = 0$ as the \emph{present} $\Present$. We call
the infinite sequence $\MeasSymbol_{-\infty : 0 }$ the \emph{past}, which we
also (more frequently) denote with an arrow pointing left: $\PastSmashed$.
Accordingly, the infinite sequence $\MeasSymbol_{1 : \infty}$ is called the
\emph{future} and denoted $\FutureSmashed$. Note that due to process
stationarity, the index denoting the present nominally can be set to any
value without altering any subsequent analysis.

Our strategy for developing the information theoretics of stochastic processes
primarily is concerned with profiling the relationships between the past,
present, and future. Given this, a useful perspective on processes is
to picture them as an communication channel transmitting information from the
past $\Past = \dots \MSym_{-3} \MSym_{-2} \MSym_{-1}$ to the future $\Future =
\MSym_1 \MSym_2 \MSym_3 \dots$ through the medium of the present $\Present$.
This perspective motivates deviating from three-way symmetry in our
i-diagrams of processes, as in \cref{fig:iDiagramProcess}. The past and the
future are depicted here as extending to the left and the right, respectively,
to mirror visualizing the bi-infinite chain of random variables. 

\subsection{Process Information Atoms}
\label{sec:ProcessInfoAtoms}

Although one might expect increasing difficulty when moving to a dynamical
system, on the surface profiling a process' information atoms in terms of
its past $\PastSmashed$, present $\Present$, and future $\FutureSmashed$
requires no more tools than already developed in \cref{sec:InformationTheory}.
We need only apply \cref{eq:infoAtomOperator} to obtain the set of appropriate
irreducible atoms:
% To simplify notation a bit, we will write the generator set of the information
% atoms of a single measurement of a process $\Process$ as $\GenSet_{\Process}$.
% In other words, we write 
\begin{align*}
    \GenSet_{\Process}
    = &  \Big\{ 
         \Hcond{\Past}{\Present, \Future} , \Hcond{\Present}{\Past, \Future} ,\\
      &  \quad \Icond{\Past}{\Present}{\Future}, \Hcond{\Future}{\Present , \Past}, \\ 
      &  \quad \Icond{\Past}{\Future}{\Present}, \Icond{\Present}{\Future}{\Past}, \\
      &  \quad \operatorname{I} \left[ \Past ; \Present ; \Future \right] \Big\} ~. 
\end{align*}

As there are only three (admittedly aggregate) random variables in play,
applying \cref{eq:infoAtomOperator} gives the expected set of seven quantities.
The atoms are shown in information diagram form in \cref{fig:iDiagramProcess}
(Below), alongside an infinite length chain (Above) depicting the measurements
of the associated process. The shape of the i-diagram has been distorted from
the symmetrical one in \cref{fig:iDiagramThreeVariable} to emphasize the
empirically known relationships between the variables (i.e., their temporal
ordering). It is worth confirming that each atom in
\cref{fig:iDiagramProcess} is identifiable as one of the atoms depicted in
\cref{fig:iDiagramThreeVariable}. 

Five out of the seven atoms in $\GenSet_{\Process}$ have been named and can be
explained intuitively \cite{Jame11a}:
\begin{enumerate}
\item \emph{Ephemeral} $\rmu$: The information localized to single measurement 
	of $\Process$ at one time and not correlated to its peers:
    \begin{align}
        \rmu =  \Hcond{\Present}{ \Past,  \Future } 
		~.
        \label{eq:ephemeralInfo}
    \end{align}
\item \emph{Binding} $\bmu$: Two equivalent quantities, \emph{forward binding
	information} $\bmuforward$ and \emph{reverse binding information}
	$\bmureverse$:
    \begin{align}
        \bmuforward & = \Icond {\Present} {\Future} {\Past} \nonumber
		~\text{and}\\ 
        \bmureverse & = \Icond {\Present} {\Past} {\Future}  
		~.
        \label{eq:bindingInfo}
    \end{align}
	For stationary processes we always have $\bmuforward = \bmureverse$. The
	forward and reverse binding informations can be interpreted as how
	correlated any given measurement of a process is with the future and the
	past, respectively. 
\item \emph{Enigmatic} $\qmu$: Aptly named, this is the interaction information
    between any given measurement of a process and the infinite past and future:
    \begin{align}
        \qmu = \operatorname{I} \left[ \Present ; \Past ; \Future \right]  
		~.
        \label{eq:EnigmaticInfo}
    \end{align}
    As this is a multivariate mutual information, it can be negative.
\item \emph{Elusive} $\sigmu$: The amount of information shared between the past
    and future that is not communicated through the present:
    \begin{align}
        \sigmu = \Icond { \Past}{ \Future}{ \Present } 
		~.
        \label{eq:ElusiveInfo}
    \end{align}
\end{enumerate}
Note that the $\measure$ in these refers to the process measure defined
in \cref{sec:Processes} and is historical notation.

The Shannon entropy rate $\hmu$ is not an irreducible information atom. It is
given by $\hmu = \Hcond{\Present}{\PastSmashed} = \bmuforward + \rmu$. As long
as the process is finitary, which is to say its \emph{excess entropy}
$\ExcessEntropy = \I{\PastSmashed}{\FutureSmashed} = \bmuforward + \qmu +
\sigmu$ is finite, the atoms above will be finite. 

The other two atoms, $\Hcond{\PastSmashed}{\Present, \FutureSmashed}$ and
$\Hcond{\FutureSmashed}{\Present, \PastSmashed}$ are typically infinite,
although they scale linearly with the length ($\ell$) of a window stretching
into the future and past: 
\begin{align*}
    \Hcond{\Past^{\ell}}{\Present, \MSym} & \sim \ell \hmu , \quad \text{and} \\
    \Hcond{\Future^{\ell}}{\Present, \MSym} & \sim \ell \hmu
	~.
\end{align*}

\section{Optimal Models of Discrete Processes}
\label{sec:CompMech}

Directly working with processes---nominally, infinite sets of infinite sequences
and their probabilities---is cumbersome. Practically, we do not want to
determine entropies over distributions of infinite pasts and futures. Rather, we
wish to build a minimal (finitely-specified) model that captures all
correlations in stochastic process $\Process$ relevant to the present
$\Present$, allowing access to a process' complete informational profile. The
framework of \emph{computational mechanics} \cite{Shal98a} provides an exact
solution to the problem of optimal minimal predictive modeling in the form of
the \emph{\eM}. We review this construction here, as well as the construction
of the reverse \eM and the bidirectional machine as introduced in Refs.
\cite{Crut08b, Jame13a}.

% \begin{definition}
%     \label{Def:HMC}
%         A finite-state edge-labeled \emph{hidden Markov model} (HMM) consists of:
%         \begin{enumerate}
%             \setlength{\topsep}{0mm}
%             \setlength{\itemsep}{0mm}
%             \item a finite set of \emph{states} $\CausalStateSet =
%                   \left\{\causalstate_1, \dots, \causalstate_N \right\}$,
%             \item a finite alphabet $\MeasAlphabet$ of $k$ \emph{symbols}
%                   $\meassymbol \in \alphabet$, and
%             \item a set of $N$ by $N$ \emph{symbol-labeled transition matrices}
%                   $T^{(\meassymbol)}$, $\meassymbol \in \alphabet$:
%                   $T^{(\meassymbol)}_{ij} = \Pr \left( \causalstate_j, \meassymbol
%                   \mid \causalstate_i \right)$.
%         \end{enumerate}
%     \end{definition}

\subsection{Computational Mechanics}
\label{sec:eMachines}

\begin{figure}
\includegraphics[width=\columnwidth]{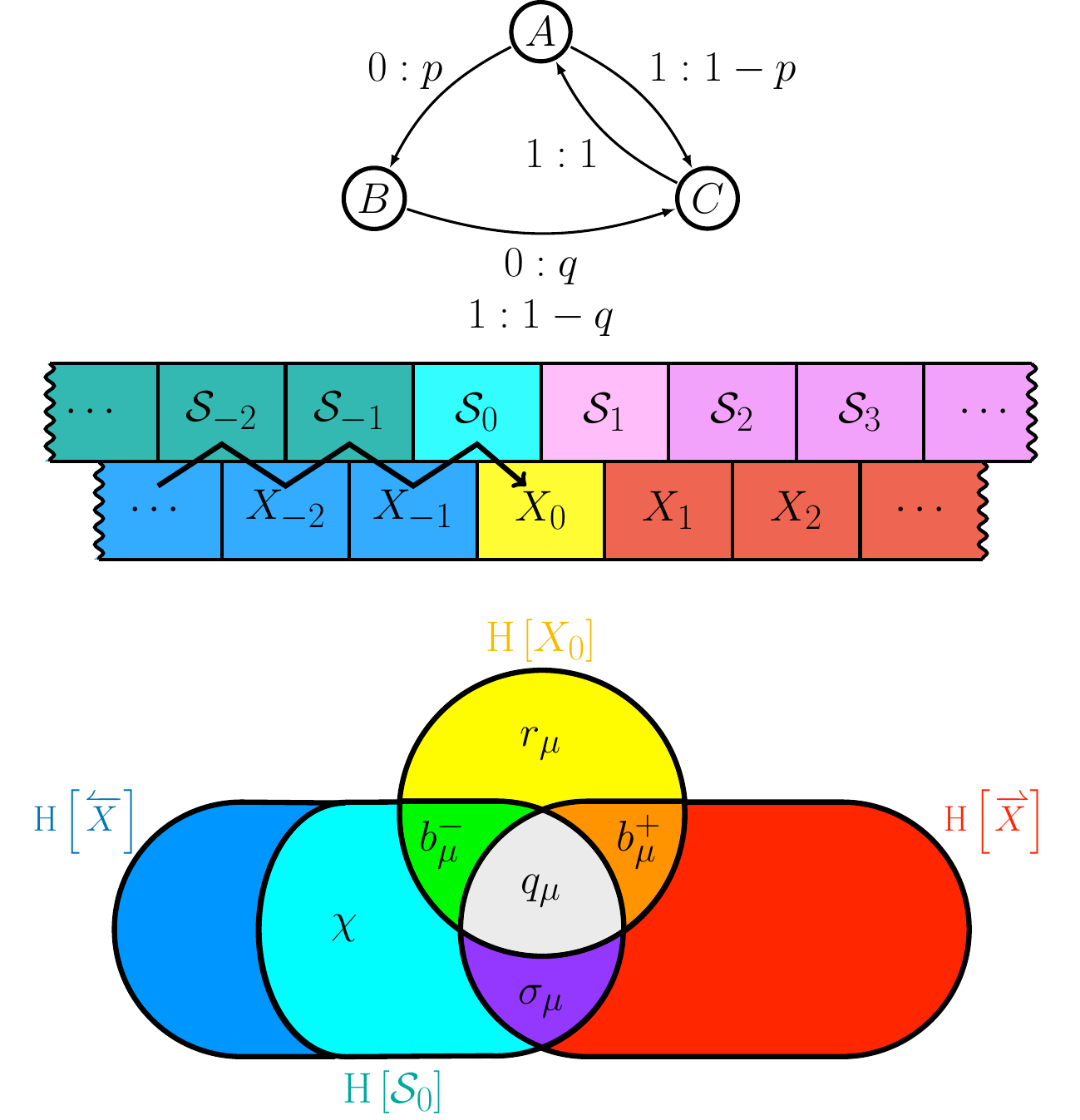}
\caption{(Top) A process' \eM as a state-transition diagram---a stochastic
	state machine. (Middle) Time indexing of causal states and measurements
	represented on an bi-infinite chain. The arrow depicts the trajectory
	(random variable sequence) through time. (Bottom) Process information
	diagram with the casual state $\CausalState_0$ at time $t=0$; cf.
	\cref{fig:iDiagramProcess}. The causal state is a function of the infinite
	past---which is to say its atom $H[\CausalState_0]$ in the i-diagram is
	contained entirely within the past $H[\Past]$.  The model complexity measure
	$\Crypticity$ (crypticity) is shown alongside the process-defined
	quantities in \cref{sec:ProcessInfoAtoms}.
	}
\label{fig:idigaramProcessS0}
\end{figure}

The states of a process' minimal optimal predictor, called the \eM, are the
classes defined by an equivalence relation $\past \sim \past '$ that groups all
pasts giving rise to the same prediction. These classes are called the
\emph{causal states}.

\begin{definition}
\label{Def:CausalStates}
A process' \emph{causal states} are the members of the range of the function:
\begin{align*}
        \EquiFunction{\past} = 
        \Big\{ \pastprime \mid
        \Pr & \left( \Future = \future | \Past = \past \right)  \\
        & = \Pr \left( \Future =  \future | \Past = \pastprime \right) \\
        & \quad \text{for all} \:\: \past \in \Past, \pastprime \in \Past \Big\}
\end{align*}
that maps from pasts to sets of pasts: $\epsilon: \AllPasts \to
\CausalStateSet$. The latter is the set of causal states, with corresponding
random variable $\CausalState$ and realizations $\causalstate$.
\end{definition}

The causal states partition the space $\AllPasts$ of all pasts into sets
(causal states $\causalstate \in \CausalStateSet$) of pasts that are
predictively equivalent. The set of causal states $\CausalStateSet$ may be
finite, fractal, or continuous, depending on the properties of the underlying
process \cite{Jurg20c}. In the following, we focus on processes with finite
causal state sets. 

The dynamic over the casual states is inherited from the shift operator
$\shiftOperator$ on the process. State-to-state transitions occur on
measurement of a new symbol $\Present = \meassymbol$, which is appended to the
observed history to give a new history: $\past \to \past \meassymbol$.
Therefore, the causal state transition is $\EquiFunction{\past} =
\causalstate_i \to \EquiFunction{\past \meassymbol} = \causalstate_j$ and
occurs with probability $\Pr \left( \Present = \meassymbol \mid \CausalState_0
= \causalstate_i \right)$. Note that the subscripts on the realizations
$\causalstate$ indicate a specific element of $\CausalStateSet$, while the
subscripts on the random variables $\MSym$ and $\CausalState$ indicate time.
\Cref{sec:tempIndexCS} discusses the temporal indexing of causal states in more
detail.

% \begin{definition}
%     \label{def:eMachine}
%     The \emph{\eM} $\eMachine$ of a finite-state process consists of 
%     \begin{enumerate}
%         \item a finite set of $\numCS$ \emph{casual states} $\CausalStateSet =
%         \left\{ \causalstate_1, \dots, \causalstate_N \right\}$,
%         \item a finite alphabet $\alphabet$ of $\numSyms$ symbols $\meassymbol \in
%         \alphabet$, and 
%         \item a set of $\numSyms$ $\numCS \times \numCS$ symbol-labeled
%             transition matrices 
%             $T^{(\meassymbol)}$, $\meassymbol \in \alphabet$: 
%             $T^{(\meassymbol)}_{ij} = \Pr \left( \causalstate_j,
%             \meassymbol \mid \causalstate_i \right)$.
%     \end{enumerate}
% \end{definition}

The causal state set together with this dynamic is the \emph{\eM} $\eMachine =
\left\{\CausalStateSet, \alphabet, \left\{T^{(\meassymbol)} : \meassymbol \in
\alphabet \right\} \right\}$, where $T^{(\meassymbol)}_{ij} = \Pr \left(
\causalstate_j, \meassymbol \mid \causalstate_i \right)$. In
\cref{fig:idigaramProcessS0} an example \eM is drawn as a state-transition
diagram with transition probabilities $\Pr\left( \MeasSymbol_t = \meassymbol
\mid \CausalState_t, \CausalState_{t+1} \right)$ from $\CausalState_t$ to
$\CausalState_{t+1}$ written as $\meassymbol : \Prob$. 

The \eM is guaranteed to be optimally predictive because knowledge of what
causal state a process is in at any time is equivalent to knowledge of the
entire past: $\Pr \left( \FutureSmashed \mid \CausalState \right) = \Pr \left(
\FutureSmashed \mid \PastSmashed \right)$. The dynamic over causal states is
Markovian in that they render the past and future statistically independent:
$\Pr \left( \PastSmashed, \FutureSmashed \mid \CausalState \right) = \Pr \left(
\PastSmashed \mid \CausalState \right) \Pr \left( \FutureSmashed \mid
\CausalState \right)$. We call these properties together \emph{causal
shielding}. \EMs also have a property called \emph{unifilarity}, which means
that knowledge of the current causal state and the next symbol is sufficient to
determine the next state: $\Hcond{\CausalState_{t+1}}{\MeasSymbol_t,
\CausalState_t} = 0$.

% This definition of the \eM identifies it as a finite-state hidden Markov model
% (HMM). Not all HMMs are \eMs and not all \eMs are HMMs, but all of the \eMs we
% will discuss in this paper are HMMs. An HMM-style \eM may be represented by a
% directed graph where the causal states are represented by the vertices and
% transitions between them by directed edges labeled with the symbol emitted on
% transition followed by the probability of transition, e.g. $\meassymbol :
% \Pr\left(\meassymbol \right)$. The time indexing is as follows: at time $t$, we
% are in state $\CausalState_t$, which emits symbol $\meassymbol_t$ and
% transitions forward to the next state $\CausalStateSet_{t+1}$. Notice that due
% to unifilarity, there is at most one transition from each causal state per
% symbol. 

% The asymptotic, stationary state distribution of the \eM is $\pi = \left\{
% \Pr(\causalstate) : \causalstate \in \CausalStateSet \right\}$ and, as a vector,
% is given by $T$'s left eigenvector corresponding to the unity eigenvalue: $\pi =
% \pi T$. 

These properties are visually represented in \cref{fig:idigaramProcessS0},
where the information $H[\CausalState_0]$ contained in causal state
$\CausalState_0$ is entirely encapsulated by the information $H[\Past]$ in the
past $\Past$. The casual state also must encompass the entirety of the excess
entropy $\ExcessEntropy = \I{\Past}{\Future}$. These two constraints result in
an i-diagram that contains strictly fewer atoms than four random variables
would maximally allow. In this case, an i-diagram has a maximum of nine random
variables. This constraint makes i-diagrams a useful tool to study \eMs beyond
the point they would normally become intractable for sets of random variables. 

The \eM is the minimal model in the sense that the amount of information stored
by the states is smaller than any other optimal rival model. We quantify this by
taking the Shannon entropy over the causal states $\Cmu = \H{\CausalState}$,
which we call the \emph{statistical complexity} \cite{Shal98a}. The difference
between model information and the excess entropy is called the \emph{crypticity}
\cite{Crut08a}:
\begin{align*}
    \Crypticity = \Cmu - \ExcessEntropy ~.
\end{align*}
$\Crypticity$ is an additional measure of model complexity that quantifies how
much internal-state information is not directly available through measurement
sequences.

\subsection{Directional Computational Mechanics}
\label{sec:directedCompMech}

While computational mechanics is built under the assumption of optimizing over
prediction, it can also be applied to the goal of \emph{retrodiction}---finding
a distribution over pasts given knowledge of the future \cite{Crut08b}. We can
think of this, equivalently, as predicting the \emph{reverse process}---the
process in a world where time runs in the opposite direction. 

\subsubsection{Reverse \texorpdfstring{$\epsilon$}{e}-Machine}
\label{sec:reverseEM}

Informationally speaking, the time-reversal of a stationary process is not
particularly interesting. As noted in \cref{sec:InfoTheoryofSP}, the forward and
reverse binding informations $\bmu$ are equal, and the excess entropy
$\ExcessEntropy$, the ephemeral information $\rmu$, the enigmatic information
$\qmu$, and the elusive information $\qmu$ are all time symmetric by definition. 

However, it is not generally the case that the predictive causal states are the
same as the retrodictive ones. And so, for a full analysis of a process'
informational structure we must consider the directional casual states. Their
construction is straightforward but requires new notation. We rename the
objects defined in \cref{Def:CausalStates} to the \emph{forward causal states}
$\forwardcausalstate \in \ForwardCausalStateSet$ and denote the equivalence
function as $\epsilon^+ \left[ \past \right]$. Similarly, the associated \eM
will now be called the \emph{forward \eM} and be denoted $\ForwardEM$. The
definitions do not change. In contrast, we have:
\begin{definition}
\label{Def:ReverseCausalStates}
A process' \emph{reverse causal states} are the members of the range of
the function:
\begin{align*}
        \epsilon^{-} \left[ \future \right] = 
        \Big\{ \futureprime \mid  & 
        \Pr \left( \Past = \past | \Future = \future \right)  \\
        = & \Pr \left( \Past =  \past | \Future = \futureprime \right) \\
        & \text{for all} \:\: \future \in \Future, \futureprime \in \Future \Big\}
\end{align*}
that maps from futures to sets of futures. The set of reverse causal states is
denoted $\ReverseCausalStateSet$, with corresponding random variable
$\ReverseCausalState$ and realizations $\reversecausalstate$. 
\end{definition}

The \emph{reverse \eM} $\ReverseEM$ is defined in the expected way, running the
shift operator $\shiftOperator$ in reverse time. It is worth noting that the
reverse \eM is not guaranteed to be finite when the forward \eM is finite, and
vice versa. However, the following will consider processes for which both
machines are finite.

As noted above, the statistical complexity $\Cmu$ typically differs in the
forward and reverse directions. Accordingly, we also have directional
crypticities with more concise expressions than those given above: 
\begin{align}
    \Crypticity^{+} & = \Hcond{\ForwardCausalState_t}{\ReverseCausalState_t}
	~\text{and} \\
    \Crypticity^{-} & = \Hcond{\ReverseCausalState_t}{\ForwardCausalState_t}
	~.
\label{eq:directionalCrypt}
\end{align}
% We call the difference the \emph{casual
% irreversibility}:
% \begin{align}
%     \Irreversibility = \ForwardCmu - \ReverseCmu ~. 
%     \label{eq:Irreversibility}
% \end{align}
% If the casual irreversibility is nonzero it implies that the forward and reverse
% crypticities differ: $\Irreversibility \neq 0 \implies \ForwardCrypticity \neq
% \ReverseCrypticity$.
The crypticities $\ForwardCrypticity$ and $\ReverseCrypticity$ have compelling
interpretations. $\ForwardCrypticity$ is the amount of information in the
forward \eM that is not contained in the excess entropy---which, recall, is the
total amount of information the process communicates through time.

It may seem odd that the causal states could contain more information than
this, but consider the classic example of a ``nearly''-IID process. Such a
process looks arbitrarily close to random, and so the amount of information
communicated through time is vanishingly small. However, in fact, there exist
very long-range correlations that can marginally improve on optimal prediction,
which must therefore be stored in the causal states. Indeed, it is not only
possible, but even typical for processes generated by hidden Markov models for
the excess entropy to be finite while the statistical complexity and therefore
the crypticity, diverge \cite{Jurg20c}. 

\subsubsection{Bidirectional Machine}
\label{sec:biMachine}

With both the forward \eM and the reverse \eM in hand, we can describe the
\emph{bidirectional machine} $\BiEM$, which simultaneously optimally predicts
and retrodicts \cite{Crut08b}.

\begin{definition}
\label{Def:BiCausalStates}
The \emph{bidirectional causal states} of a process are the members of the
range of the function:
\begin{align*}
        \epsilon^{\pm} \left[ \biinfinity = \left( \past, \future \right) \right] = 
        \Big\{  \left( \pastprime, \futureprime \right) \mid  
        % & \Pr \left( \Future = \future | \Past = \past \right)  \\
        % = & \Pr \left( \Future =  \future | \Past = \pastprime \right) \\
        % & \text{for all} \:\: \past \in \Past, \pastprime \in \Past \text{; and } \\
        % & \Pr \left( \Past = \past | \Future = \future \right)  \\
        % = & \Pr \left( \Past =  \past | \Future = \futureprime \right) \\
        % & \text{for all} \:\: \future \in \Future, \futureprime \in \Future \Big\}
        & \pastprime \in \epsilon^+ \left[ \past \right] \: \text{and} \\ 
        & \futureprime \in \epsilon^- \left[ \future \right]  \Big\}
\end{align*}
that maps histories to sets of histories. The set of bidirectional causal
states is denoted $\BiCausalStateSet$, with corresponding random variable
$\BiCausalState$ and realizations $\bicausalstate$. 
\end{definition}

The bidirectional causal states are a subset of the Cartesian product of forward
and reverse casual states: $\BiCausalStateSet \subseteq \ForwardCausalStateSet
\times \ReverseCausalStateSet$. Our convention in the following is to label
causal states with Latin letters, using upper case for the forward direction and
lower case for the reverse direction: i.e., $\ForwardCausalStateSet = \left\{ B
C \right\}$ and $\ReverseCausalStateSet = \left\{ a, b, c, d \right\}$ as in
\cref{fig:emachine_ordering}. The bidirectional states are labeled by their
corresponding forward and reverse states: i.e., $\BiCausalStateSet = \left\{ Aa,
Ba, \dots \right\}$. Transition labels are written to indicate the direction:
$\meassymbol : \Prob : \text{direction}$. See \cref{fig:iDiagramseMachines} for
examples. 

We primarily use the bidirectional machine in the algorithm that calculates
our new informational properties, as discussed in \cref{sec:Algorithm}.

% The bidirectional \eM $\BiEM$ has transitions in
% both the forward and reverse time directions. These edges are labeled
% $\meassymbol : \Pr \left( \meassymbol \right) : \pm$, as in
% \cref{fig:bidirectionalMachine} (c). 

% The bidirectional statistical complexity
% $\BiCmu$ is the state information required to optimally predict and retrodict
% and has a number of useful relationships: 

% \begin{align*}
%     \BiCmu = & \H{\BiCausalState} \\
%     = & \H{\ForwardCausalState, \ReverseCausalState} \\
%     = & \, \ExcessEntropy + \Crypticity \\ 
%     = & \,  \ForwardCmu + \ReverseCmu - \ExcessEntropy \\ 
%     & \leq \ForwardCmu + \ReverseCmu ~. 
% \end{align*}

% Notice that the difference between the bidirectional statistical complexity and
% the sum of the directional causal complexity is exactly the excess entropy. 

\subsubsection{Temporal Indexing of Causal States}
\label{sec:tempIndexCS}

\begin{figure}
\centering
\includegraphics[width=\columnwidth]{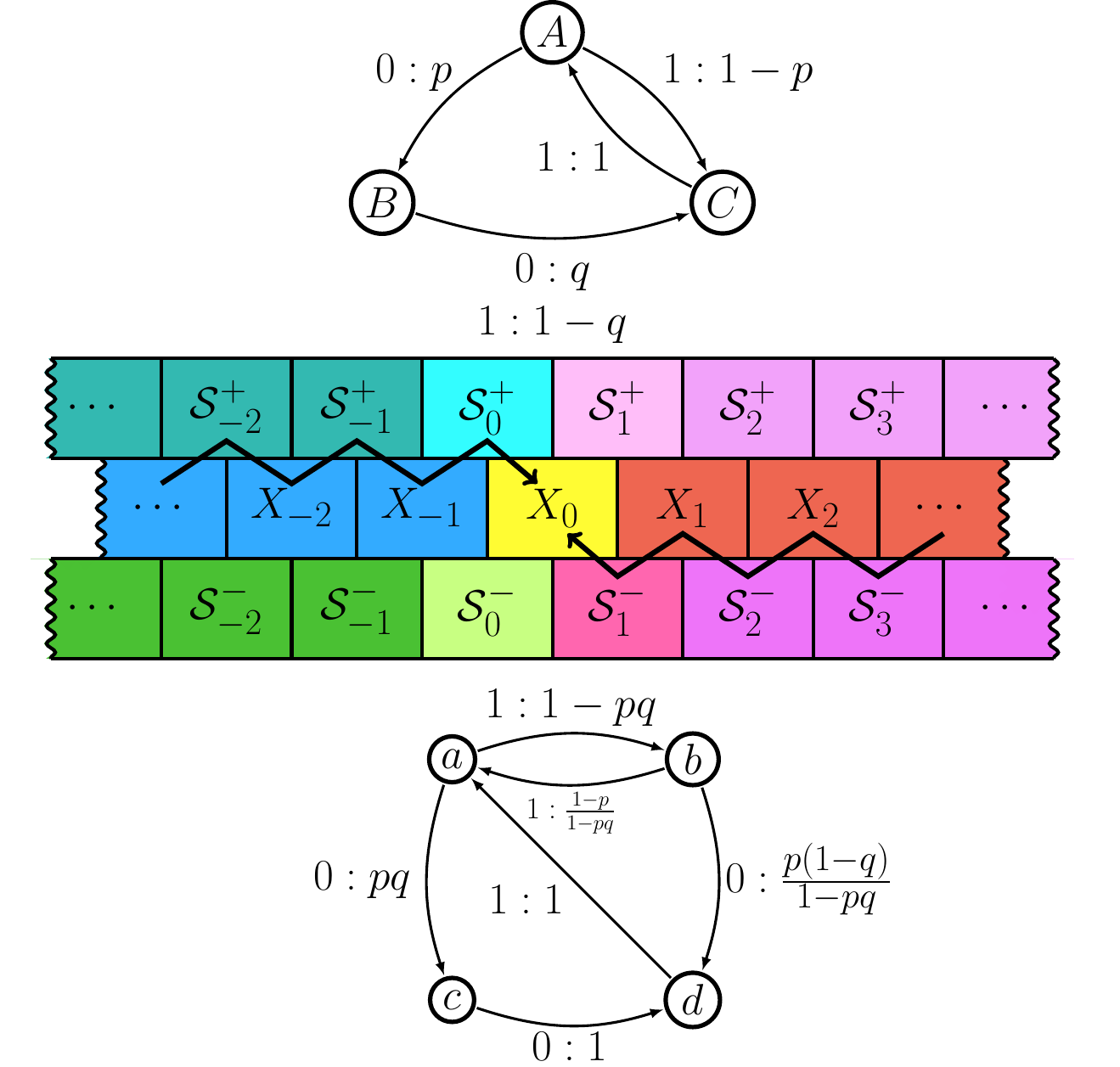}
\caption{The forward (Top) and reverse (Bottom) \eMs of a stochastic process,
	where $\CausalStateSet^+ = \{A,B,C\}$ and $\CausalStateSet^- = \{a,b,c,d\}$.
	The time indexing of the causal states and the emitted measurements are laid
	out on three parallel horizontal chains. The variables on the chain are
	color coded to match \cref{fig:iDiagramProcessS0S0S1S1}, which depicts the
	accompanying information diagram. The arrows depict the path through time in
	the forward (Top) and reverse (Bottom) directions, respectively; cf.
	\cref{fig:idigaramProcessS0}.
	}
\label{fig:emachine_ordering}
\end{figure}

% For process random variables---e.g., $\MSym$ and $\CausalState$---subscripts
% indicate the variable's time index. While subscripts on realizations indicate
% their set index. In other words, $\ForwardCausalState_t = \causalstate_i$
% refers to the forward causal state at time $t$, which takes on the value of the
% $i$th element of $\ForwardCausalStateSet$.  

\Cref{fig:emachine_ordering} depicts the forward \eM (Top) and the reverse \eM
(Bottom) of a given process. The time-indexed states of the \eMs are depicted
on state chains $\ldots \CausalState_1 \CausalState_2 \ldots$ sandwiching the
chain of process measurements $\ldots \MSym_1 \MSym_2 \ldots$. Although we
index the causal states with integers, we imagine them as occurring on ``half
time steps'' in between the measurement time indices. The arrows trace the path
through time along the causal states and observed measurements. Note that in
the forward direction, the causal state at time $t$ emits the measurement at
time $t$, but in the reverse direction the causal state at time $t$ is said to
emit the measurement at time $t-1$. This offset is a consequence of using
integer indices for the states. The mismatch in the reverse time direction
(rather than the forward direction) is a matter of convention. 

\begin{figure*}
\centering
\includegraphics[width=0.85\textwidth]{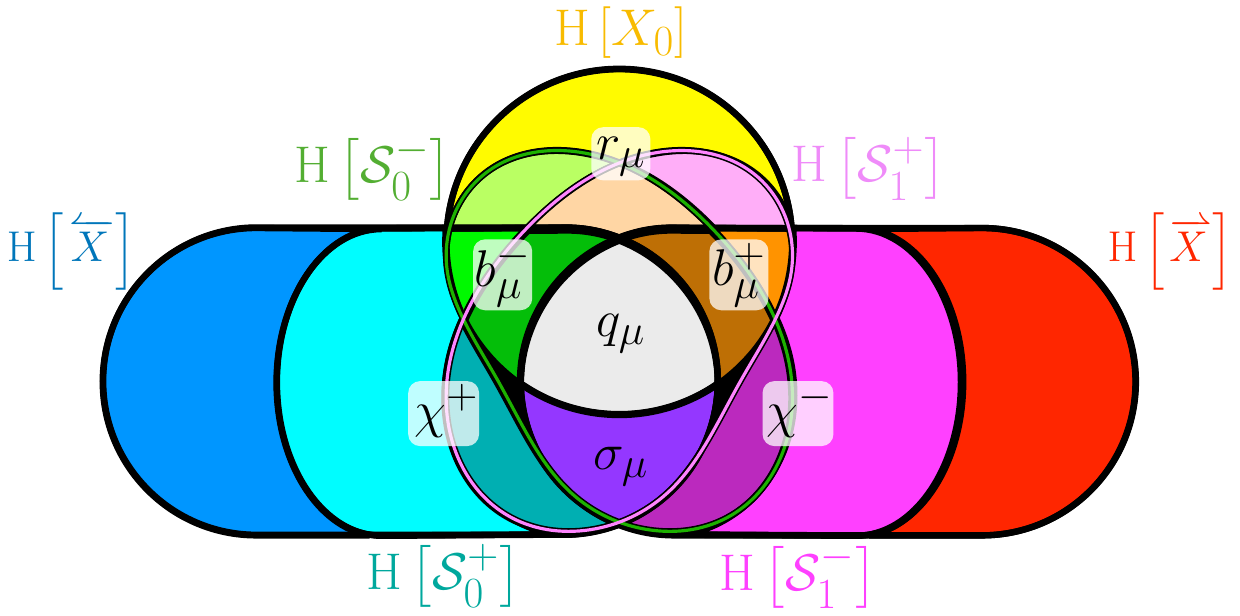}
\caption{Information diagram representing all possible positive atoms of a
    single transition of a bidirectional machine, including the states at $t=0$
    and the states at $t=1$. The majority of the information atoms theoretically
    possible go to zero due to the causal shielding of the causal states. The
    information atoms defined in \cref{sec:ProcessInfoAtoms}, five of which are
    no longer irreducible, are overlaid on their corresponding atoms.
	}
\label{fig:iDiagramProcessS0S0S1S1}
\end{figure*}

Note that there are four states that symmetrically ``surround'' each
measurement. For the present $\Present$, these states are
$\ForwardCausalState_0$, $\ReverseCausalState_0$, $\ForwardCausalState_1$, and
$\ReverseCausalState_1$. The informational relationship the forward and reverse
states have with the measurement they surround is asymmetrical. We might say
that two of the states---$\ForwardCausalState_1$ and
$\ReverseCausalState_0$---have already ``seen'' the measurement $\Present$, as
it was emitted on the transition \emph{to} that state. From the perspective of
these states, $\Present$ is included in the past or future, respectively. We say
that $\ReverseCausalState_0$ and $\ForwardCausalState_1$ are ``interior'' to the
measurement, drawing on the visual depiction in the i-diagram in
\cref{fig:iDiagramProcessS0S0S1S1}, where these states (kidney bean in shape)
are positioned as closer to the center of the diagram. The other states
$\ForwardCausalState_0$ and $\ReverseCausalState_1$ are then ``exterior''---they
trail on either end of the i-diagram due to their access to information furthest
in the past or future, respectively. 

% \subsection{Information in the Causal States}
% \label{sec:InformationinEMs}

% The \eM is the minimal model in the sense that the amount of information stored
% by the states is smaller than any other optimal rival. We quantify this by
% taking the Shannon entropy over the causal states $\Cmu = \H{\CausalState}$,
% which we call the \emph{statistical complexity}. 

\section{Atomic Taxonomy}
\label{sec:anatomyofamodel}

With the causal states in place, we can develop a full information-theoretic
analysis of prediction and retrodiction. 

\subsection{Information Atoms from Causal States}
\label{sec:infoAtomModel}

Naively, our new information atom set is formed by simply adding the four
causal states ``surrounding'' the present measurement to our random variable
set:
\begin{align*}
\RVSet_{\epsilon} = \left\{ \Past, \ForwardCausalState_0,
\ReverseCausalState_0, \Present, \ForwardCausalState_1, \ReverseCausalState_1,
\Future \right\}
~.
\end{align*}
However, thanks to causal shielding, we can drop the infinite past and future,
as they are redundant with the causal states. So our relevant random
variable set is:
\begin{align*}
\RVSet_{\epsilon} = \left\{
\ForwardCausalState_0, \ReverseCausalState_0, \Present, \ForwardCausalState_1,
\ReverseCausalState_1 \right\}
  ~.
\end{align*}

Five random variables maximally produces an irreducible atom set of $2^5 = 32$
atoms, but $\GenSet_{\BiEM}$ consists of only fourteen nonzero irreducible
atoms. This reduction is due to the particular properties of the causal
states---namely unifilarity and causal shielding. The structured nature of
\cref{fig:iDiagramProcessS0S0S1S1} indicates the influence of these properties,
which we discuss in further depth in \cref{sec:StructureineMInfoAtoms}.  First,
to get there we introduce the nonzero information atoms of an optimally modeled
process.

% \begin{align*}
% \left\{ \ForwardCausalState_0, \ReverseCausalState_0, \Present,
% \ForwardCausalState_1, \ReverseCausalState_1 \right\}
%   ~.
% \end{align*}

% \subsubsection{Casual Shielding}
% \label{sec:causalShielding}

% Our secondary purpose of introducing computational mechanics was to reap the
% benefits of the causal shielding properties of the causal states. By using the
% \eMs as our model, we can study all temporal correlations that impact the
% present without including the infinite-length past and future in our
% informational analysis, as we did in \cref{sec:ProcessInfoAtoms}. In practice,
% this means replacing the conditioning variables according to shielding order.

% Given a set of variables $\RVSet$ and subsets $A, B \in \RVSet$ we say that
% $A$ \emph{shields} the set $\RVSet$ from $B$ if:
% \begin{align*}
%     \left( C \perp B  \mid A \right) ~,
% 	~\text{for~all~} C \in (\RVSet \setminus B)
% 	~. 
% \end{align*}

% For example, $\ForwardCausalState_0$ shields $\RVSet_{\epsilon}$ from the past
% $\Past$ and $\ReverseCausalState_1$ shields $\RVSet_{\epsilon}$ from the
% future $\Future$. This means that wherever $\Past$ appears in an informational
% quantity, it is possible to replace the variable with $\ForwardCausalState_0$
% with no loss of information and to replace $\Future$ with $\ReverseCausalState_1$.

\subsubsection{Anatomy of a Bit Redux}
\label{sec:anatomyofaBit}

Ten of our new information atoms are related to the original five atoms given
in \cref{sec:ProcessInfoAtoms}. First, rewrite those atoms in terms of the
causal states, replacing infinite futures and pasts with the appropriate
shielding causal states: 
\begin{itemize}
    \item $\rmu = 
    \Hcond{\Present}{\ForwardCausalState_0, \ReverseCausalState_1}$ ~,
    \item $\bmuforward = 
    \Icond{\Present}{\ReverseCausalState_1}{\ForwardCausalState_0}$ ~,
    \item $\bmureverse = 
    \Icond{\Present}{\ForwardCausalState_0}{\ReverseCausalState_1}$ ~,
    \item $\qmu = 
    \operatorname{I} \left[ \ForwardCausalState_0 ; \Present ;
    \ReverseCausalState_1 \right]$ ~, ~and
    \item $\sigmu = 
    \Icond{\ForwardCausalState_0}{\ReverseCausalState_1}{\Present}$
	~.
\end{itemize}
The increase in number of atoms from five to ten is due to the ``splitting'' of
the binding informations $\bmu$ and the ephemeral information $\rmu$ into
\typeOne and \typeTwo pieces.

By \emph{\typeOne information} we refer to information that will be
``forgotten'' by the the \eMs within a single time step, either into the future
(for the forward \eM) or into the past (for the reverse \eM). By \emph{\typeTwo
information} we mean information that is ``stored'' in the model, and remains
correlated with new causal states.

\Cref{fig:iDiagramProcessS0S0S1S1} depicts this by overlaying the taxonomy of a
process' informational quantities over their new constituent atoms. The
\typeTwo informations are colored darker in shade. The full list of atoms is
given by \cref{tab:InfoAtomsModel}, organized by their parent ``anatomy of a
bit'' quantity.

\begin{figure}
\centering
\includegraphics[width=0.4\textwidth]{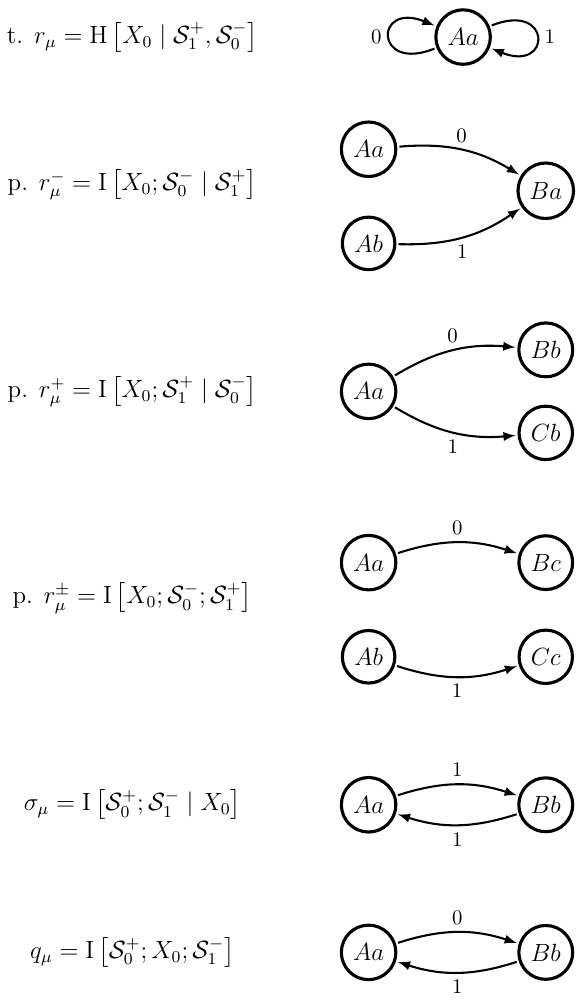}
\caption{Machine (state-transition) ``motifs'' underlying the ephemeral
	informations, the enigmatic information, and the elusive information. These
	motifs give rise only to this type of information, but the motifs are not
	necessarily the only way in which these informations may be produced. Note
	that the first, fifth, and sixth examples can be complete state machines,
	although only the first and sixth examples are \eMs. (The fifth example is nonminimal.)
    }
\label{fig:modelMotifs}
\end{figure}

Several atoms can be mapped directly to topological \emph{motifs} that capture
the time-local state transition structure that gives rise to them in the \eMs.
These atoms with their corresponding motifs are listed in
\cref{fig:modelMotifs}. The motifs make explicit the mechanism producing the
associated information measure.

The ephemeral information splits into four terms: 
\begin{align*}
\rmu  = & \text{ t. } \rmu + \text{ p. } \rmu^- + \text{ p. } \rmu^-
        + \text{ p. } \rmu^{\pm} \\
      = & \overbrace{\Hcond{\Present}
        {\ForwardCausalState_1, \ReverseCausalState_0}}^{\text{\typeOne}} \\
        & \quad + \Icond{\Present}{\ReverseCausalState_0}{\ForwardCausalState_1} 
        + \Icond{\Present}{\ForwardCausalState_1}{\ReverseCausalState_0} \\
        & \quad \underbrace{ + \, \Icond{\Present}
        {\ForwardCausalState_1; \ReverseCausalState_0}
        {\ForwardCausalState_0, \ReverseCausalState_1}  
        \hspace{4.3em}}_{\text{\typeTwo}}
   ~. 
\end{align*}

It helps to compare the terms above to the atoms of
\cref{fig:iDiagramProcessS0S0S1S1}. The first term is the \typeOne ephemeral
information, which is truly ephemeral in that it remains uncorrelated with any
causal state at any time. The remaining three are all \typeTwo: the second term
is ephemeral information that is correlated with only the states of the reverse
\eM, the third term only with states of the forward \eM, and the fourth
information term is correlated with both. 

These ephemeral quantities are produced by specific motifs in the bidirectional
machine, as shown in \cref{fig:modelMotifs}.  

Now consider first the reverse binding information $\bmureverse$. This splits
into two terms:
\begin{align*}
    \bmureverse = & \hspace{2.95em} \text{ t. } \bmureverse 
    \hspace{3em} +  \hspace{3em} \text{ p. } \bmureverse \\
	= & \underbrace{\Icond{\Present}{\ForwardCausalState_0;
	\ReverseCausalState_0}{\ForwardCausalState_1}}_{\text{\typeOne}}
	+ \underbrace{\Icond{\Present}{\ForwardCausalState_0 ; 
    \ReverseCausalState_0;\ForwardCausalState_1}
    {\ReverseCausalState_1}}_{\text{\typeTwo}}
	~.
\end{align*}
The first term is \emph{\typeOne binding information} in the forward causal
state at $t=0$ that is not carried through to the forward causal state at
$t=1$. The second term is called \emph{\typeTwo} as it is that part of the
binding information correlated with $\ForwardCausalState_1$. It therefore
influences the future states of the forward \eM. 

We can do the same analysis with the forward binding information and the
reverse causal states, recalling that the reverse \eM runs in reverse time: 
\begin{align*}
\bmuforward = & \hspace{2.95em} \text{ t. } \bmuforward 
    \hspace{3em} +  \hspace{3em} \text{ p. } \bmuforward \\
    = & \underbrace{\Icond{\Present}{\ReverseCausalState_1;
    \ForwardCausalState_1}
    {\ReverseCausalState_0}}_{\text{\typeOne}}
	+ 
    \underbrace{\Icond{\Present}{\ReverseCausalState; 
    \ReverseCausalState_0; \ForwardCausalState_1}
    {\ForwardCausalState_0}}_{\text{\typeTwo}}
	~. 
\end{align*}

The second term is \typeTwo reverse binding information correlated with
$\ReverseCausalState_0$ and it, therefore, influences \emph{past} states of the
reverse \eM. Unfortunately, these quantities do not seem to map easily to
isolated motifs in the bidirectional machine, but they do appear in the examples
we consider in \cref{sec:examples}. 

The enigmatic information $\qmu$ and elusive information $\sigmu$ are not
impacted by the addition of the causal states into our informational analysis
except to update their definitions as above. However, they are listed in
\cref{fig:modelMotifs}. Note that \typeOne $\rmu$, $\qmu$, and $\sigmu$ all
correspond to motifs that can be complete machines in and of themselves,
although $\sigmu$ is not minimal and therefore would not be an \eM. Typically,
though, we encounter these motifs as components of larger and more complex state
machines. 

\subsubsection{Splitting Causal State Information}
\label{sec:splittingModel}

We have now accounted for ten information atoms corresponding to process
measurements. There are still four purely causal model information atoms, two
of which are new to this analysis. Recall the forward and reverse crypticities
\cref{eq:directionalCrypt}. For our system, we have:
\begin{align*}
\ForwardCrypticity & = \Hcond{\ForwardCausalState_0}{\ReverseCausalState_0} \\
    \ReverseCrypticity & = \Hcond{\ReverseCausalState_1}{\ForwardCausalState_1} ~.
\end{align*}

As already noted by
\cref{sec:directedCompMech}, the crypticities are a type of modeling
information---the amount of information required for the causal states to do
optimal prediction or retrodiction above and beyond the excess entropy. As with
the binding and ephemeral informations, some of this information is \typeOne
and some \typeTwo.

Consider the forward crypticity:
\begin{align*}
    \ForwardCrypticity =  & \hspace{2.5em} \text{ t. } \ForwardCrypticity 
                \hspace{1.95em} +  \hspace{2.5em} \text{ p. } \ForwardCrypticity \\ 
            = & \underbrace{\Hcond{\ForwardCausalState_0}{\ForwardCausalState_1, 
                \ReverseCausalState_0}}_{\text{\typeOne}} + 
                \underbrace{\Icond{\ForwardCausalState_0}{\ForwardCausalState_1}
                {\ReverseCausalState_0}}_{\text{\typeTwo}} ~. 
\end{align*}
The first term is the \typeOne forward crypticity. This is modeling information
that is ``forgotten'' after one time step---necessary for optimal prediction of
$\Present$ but not of $\MSym_1$. The second term is the \typeTwo forward
crypticity, which is correlated with $\ForwardCausalState_1$ and continues to be
influential in prediction of future observations.

The reverse crypticity splits in the same manner: 
\begin{align*}
\ReverseCrypticity = & \hspace{2.5em} \text{ t. } \ReverseCrypticity 
                \hspace{1.95em} +  \hspace{2.5em} \text{ p. } \ReverseCrypticity \\ 
                = &\underbrace{\Hcond{\ReverseCausalState_1}{\ForwardCausalState_1, 
                         \ReverseCausalState_0}}_{\text{\typeOne}} + 
                         \underbrace{\Icond{\ReverseCausalState_1}{\ReverseCausalState_0}
                         {\ForwardCausalState_1}}_{\text{\typeTwo}}
	~. 
\end{align*}
Again, the first term is \typeOne and the second is \typeTwo, although in
this direction the difference is whether or not the information is correlated
with the reverse causal state $\ReverseCausalState_0$. 

\begin{table*}[]
    \def\arraystretch{1.4}
    \begin{tabular}{c
        @{\hskip 0.1in}c@{\hskip 0.1in}
        c
        @{\hskip 0.15in}rl@{\hskip 0.15in}
        D{|}{\mid}{-1}@{\hskip -4em}c}
    \toprule
    \multicolumn{3}{c}{Label Type} &
    \multicolumn{2}{c}{Partition} &
    \multicolumn{2}{c}{Information Atom} \\
    \toprule
    Decimal & 
    \multicolumn{1}{c}{Lexicographic} & 
    Indicial  &
    Joint Set &
    Conditioned &
    \multicolumn{1}{c}{Atom} &
    \multicolumn{1}{c}{Type} \\
    $i$ & 
    $\ForwardCausalState_0 \hspace{0.15em} 
    \ReverseCausalState_0 \hspace{0.15em}
    \Present \hspace{0.15em}
    \ForwardCausalState_1 \hspace{0.15em}
    \ReverseCausalState_1 $ & 
    $k$ &
    $\Asubset_i$ &
    $ \overline{\Asubset_i}$ & 
    \multicolumn{1}{c}{$\atom_i$}  & \\
    \midrule
    1 & 1 $\hspace{0.15em}$ 0 $\hspace{0.15em}$ 0 
    $\hspace{0.15em}$ 0 $\hspace{0.15em}$ 0 & 0 &
    $ \left\{ \ForwardCausalState_0  \right\} $ &
    $ \left\{ \ReverseCausalState_0, \Present, 
              \ForwardCausalState_1, \ReverseCausalState_1  \right\} $  & 
    \operatorname{H} \left[\ForwardCausalState_0 \right. | 
    \left. \ReverseCausalState_0, \ForwardCausalState_1 \right] &
    t. $\ForwardCrypticity$
    \\ 
    9 & 1 $\hspace{0.15em}$ 0 $\hspace{0.15em}$ 0 
    $\hspace{0.15em}$ 1 $\hspace{0.15em}$ 0 & 03 &
    $ \left\{ \ForwardCausalState_0, \ForwardCausalState_1  \right\} $ &
    $ \left\{ \ReverseCausalState_0, \Present, 
              \ReverseCausalState_1  \right\} $  & 
    \operatorname{I} \left[\ForwardCausalState_0; \ForwardCausalState_1 \right. | 
    \left. \ReverseCausalState_0\right] &
    p. $\ForwardCrypticity$
    \\
    & & & & & & \\ 
    7 & 1 $\hspace{0.15em}$ 1 $\hspace{0.15em}$ 1 
    $\hspace{0.15em}$ 0 $\hspace{0.15em}$ 0 & 012 &
    $ \left\{ \ForwardCausalState_0, \ReverseCausalState_0, \Present  \right\} $ &
    $ \left\{ \ForwardCausalState_1, \ReverseCausalState_1  \right\} $  & 
    \operatorname{I} \left[\ForwardCausalState_0; 
    \ReverseCausalState_0; \Present \right. | 
    \left. \ForwardCausalState_1 \right] &
    t. $\bmureverse$
    \\ 
    15 & 1 $\hspace{0.15em}$ 1 $\hspace{0.15em}$ 1 
    $\hspace{0.15em}$ 1 $\hspace{0.15em}$ 0 & 0123 &
    $ \left\{ \ForwardCausalState_0, \ReverseCausalState_0, 
              \Present, \ForwardCausalState_1  \right\} $ &
    $ \left\{ \ReverseCausalState_1  \right\} $  & 
    \operatorname{I} \left[\ForwardCausalState_0; \ReverseCausalState_0; 
    \Present; \ForwardCausalState_1 \right. | 
    \left. \ReverseCausalState_1 \right] &
    p. $\bmureverse$
    \\ 
    & & & & & & \\ 
    4 & 0 $\hspace{0.15em}$ 0 $\hspace{0.15em}$ 1 
    $\hspace{0.15em}$ 0 $\hspace{0.15em}$ 0 & 2 &
    $ \left\{ \Present  \right\} $ &
    $ \left\{ \ForwardCausalState_0, \ReverseCausalState_0, 
    \ForwardCausalState_1, \ReverseCausalState_1  \right\} $  & 
    \operatorname{H} \left[\Present \right. | 
    \left. \ReverseCausalState_0, \ForwardCausalState_1 \right] & 
    t. $\rmu$
    \\ 
    6 & 0 $\hspace{0.15em}$ 1 $\hspace{0.15em}$ 1 
    $\hspace{0.15em}$ 0 $\hspace{0.15em}$ 0 & 12 &
    $ \left\{ \ReverseCausalState_0, \Present  \right\} $ &
    $ \left\{ \ForwardCausalState_0, \ForwardCausalState_1, 
              \ReverseCausalState_1  \right\} $  & 
    \operatorname{I} \left[\ReverseCausalState_0; \Present \right. | 
    \left. \ForwardCausalState_0, \ForwardCausalState_1
    \right] &
    p. $\rmu^-$
    \\ 
    12 & 0 $\hspace{0.15em}$ 0 $\hspace{0.15em}$ 1 
    $\hspace{0.15em}$ 1 $\hspace{0.15em}$ 0 & 23 &
    $ \left\{ \Present, \ForwardCausalState_1  \right\} $ &
    $ \left\{ \ForwardCausalState_0, \ReverseCausalState_0, 
              \ReverseCausalState_1  \right\} $  & 
    \operatorname{I} \left[\Present; \ForwardCausalState_1 \right. | 
    \left. \ReverseCausalState_0, \ReverseCausalState_1
    \right] &
    p. $\rmu^+$
    \\ 
    14 & 0 $\hspace{0.15em}$ 1 $\hspace{0.15em}$ 1 
    $\hspace{0.15em}$ 1 $\hspace{0.15em}$ 0 & 123 &
    $ \left\{ \ReverseCausalState_0, \Present, 
            \ForwardCausalState_1  \right\} $ &
    $ \left\{ \ForwardCausalState_0, \ReverseCausalState_1  \right\} $  & 
    \operatorname{I} \left[\ReverseCausalState_0; 
    \Present; \ForwardCausalState_1 \right. | 
    \left. \ForwardCausalState_0, \ReverseCausalState_1 \right] &
    p. $\rmu^{\pm}$
    \\ 
    & & & & & & \\ 
    28 & 0 $\hspace{0.15em}$ 0 $\hspace{0.15em}$ 1 
    $\hspace{0.15em}$ 1 $\hspace{0.15em}$ 1 & 234 &
    $ \left\{ \Present, \ForwardCausalState_1, 
              \ReverseCausalState_1  \right\} $ &
    $ \left\{ \ForwardCausalState_0, \ReverseCausalState_0  \right\} $  & 
    \operatorname{I} \left[\Present; \ForwardCausalState_1; 
    \ReverseCausalState_1 \right. | 
    \left. \ReverseCausalState_0 \right] &
    t. $\bmuforward$
    \\ 
    30 & 0 $\hspace{0.15em}$ 1 $\hspace{0.15em}$ 1 
    $\hspace{0.15em}$ 1 $\hspace{0.15em}$ 1 & 1234 &
    $ \left\{ \ReverseCausalState_0, \Present, \ForwardCausalState_1, 
              \ReverseCausalState_1  \right\} $ &
    $ \left\{ \ForwardCausalState_0  \right\} $  & 
    \operatorname{I} \left[\ReverseCausalState_0; \Present; 
    \ForwardCausalState_1; \ReverseCausalState_1 \right. | 
    \left. \ForwardCausalState_0 \right] &
    p. $\bmuforward$
    \\ 
    & & & & & & \\ 
    16 & 0 $\hspace{0.15em}$ 0 $\hspace{0.15em}$ 0 
    $\hspace{0.15em}$ 0 $\hspace{0.15em}$ 1 & 4 &
    $ \left\{ \ReverseCausalState_1  \right\} $ &
    $ \left\{ \ForwardCausalState_0, \ReverseCausalState_0, 
              \Present, \ForwardCausalState_1  \right\} $  & 
    \operatorname{H} \left[\ReverseCausalState_1 \right. | 
    \left.\ReverseCausalState_0, \ForwardCausalState_1 \right] &
    t. $\ReverseCrypticity$
    \\ 
    18 & 0 $\hspace{0.15em}$ 1 $\hspace{0.15em}$ 0 
    $\hspace{0.15em}$ 0 $\hspace{0.15em}$ 1 & 14 &
    $ \left\{ \ReverseCausalState_0, \ReverseCausalState_1  \right\} $ &
    $ \left\{ \ForwardCausalState_0, \Present, 
              \ForwardCausalState_1  \right\} $  & 
    \operatorname{I} \left[\ReverseCausalState_0; 
    \ReverseCausalState_1 \right. | 
    \left. \ForwardCausalState_1 \right] &
    p. $\ReverseCrypticity$
    \\ 
    & & & & & & \\ 
    27 & 1 $\hspace{0.15em}$ 1 $\hspace{0.15em}$ 0 
    $\hspace{0.15em}$ 1 $\hspace{0.15em}$ 1 & 0134 &
    $ \left\{ \ForwardCausalState_0, \ReverseCausalState_0, 
              \ForwardCausalState_1, \ReverseCausalState_1  \right\} $ &
    $ \left\{ \Present  \right\} $  & 
    \operatorname{I} \left[\ForwardCausalState_0; \ReverseCausalState_0; 
    \ForwardCausalState_1; \ReverseCausalState_1 \right. | 
    \left. \Present \right] &
    $\sigmu$
    \\ 
    & & & & & & \\ 
    31 & 1 $\hspace{0.15em}$ 1 $\hspace{0.15em}$ 1 
    $\hspace{0.15em}$ 1 $\hspace{0.15em}$ 1 & 01234 &
    $ \left\{ \ForwardCausalState_0, \ReverseCausalState_0, \Present, 
              \ForwardCausalState_1, \ReverseCausalState_1  \right\} $ &
    $ \emptyset $  & 
    \operatorname{I} \left[\ForwardCausalState_0; \ReverseCausalState_0; 
    \Present; \ForwardCausalState_1; \ReverseCausalState_1 \right] &
    $\qmu$
    \\    
    \bottomrule
    \end{tabular}
\caption{Irreducible, nonzero information atoms for five random variables
	$\RVSet = \left\{\ForwardCausalState_0; \ReverseCausalState_0; \Present;
	\ForwardCausalState_1; \ReverseCausalState_1 \right\}$ for a given process.
	The decimal, lexicographic, and indicial labels are given in the left side
	columns, as laid out in \cref{sec:infoatomLabeling}. The partitioning of
	the variables is given in the middle two columns, with variables in the
	left side in the joint distribution and variables on the right side in the
	conditioning distribution. On the far right, the corresponding information
	atom is written explicitly (with redundant conditioning variables dropped)
	alongside the ``type'' of atom in the taxonomic scheme given in
	\cref{sec:ProcessInfoAtoms} and whether it is \typeOne (t.) or \typeTwo (p.).
	}
\label{tab:InfoAtomsModel}
\end{table*}

\subsection{Atomic Indicial Structure}
\label{sec:StructureineMInfoAtoms}

As already noted, our informational taxonomy of a prediction results in only
fourteen atoms despite a theoretically-possible set of thirty two. This
reduction is a result of the structural properties of the causal states. These
properties are concisely described using the indicial labeling described in
\cref{sec:infoatomLabeling}. Our convention is to order sequences of causal
states and measurements starting with a forward-time causal state and continuing
in the order: $\ForwardCausalState_t, \ReverseCausalState_t, \MeasSymbol_t,
\ForwardCausalState_{t+1}, \ReverseCausalState_{t+1}, \MeasSymbol_{t+1}, \dots$.

This means that in the indicial notation, we have: 
\begin{align*}
    \ForwardCausalState_t \hspace{.5em} \to & \hspace{1em} k = t \\ 
    \ReverseCausalState_t \hspace{.5em} \to &\hspace{1em}  k = t + 1 \\ 
    \MeasSymbol_t \hspace{.5em} \to & \hspace{1em} k = t + 2 
	~.
\end{align*}

Using the shorthand notation $\H{k} = \H{\ForwardCausalState_t}$, we can
then express the structural properties in terms of patterns in the indexes of
the random variables, as follows:  
\begin{enumerate}
\item \emph{Unifilarity}: Given a measurement and the causal state that emitted
	it, there is no longer any uncertainty in the next state. In the forward and
    reverse directions, for $k \in \nats, k \bmod 3 = 0$, the disallowed atoms
    are given by: 
    \begin{align*}
        \Hcond{k+3; \dots}{k, k+2, \dots} = & \, 0   ~\text{and}\\
        \Hcond{k+1; \dots}{k+2, k+4, \dots} = & \, 0
		~,
    \end{align*}
    where the dots indicate that the remaining two variables may be added to
    either side of the partition. For our analysis of the present, this zeroes
    out four atoms in each direction.
    % \begin{align*}
    % \def\arraystretch{1.4}
    % \begin{tabular}{D{|}{\mid \,}{-1}@
    %                 {\hskip 2em}
    %                 D{|}{\mid \,}{-1}}
    %     \multicolumn{1}{c}{Forward} & \multicolumn{1}{c}{Reverse} \\
    %     \midrule
    %     \operatorname{H} \left[ \, 3 \right. | 0124 \left. \right] & 
    %     \operatorname{H} \left[ \, 1 \right. | 0234 \left. \right] \\
    %     \operatorname{I} \left[ \, 13 \right. | 024 \left. \right] & 
    %     \operatorname{I} \left[ \, 01 \right. | 234 \left. \right] \\
    %     \operatorname{I} \left[ \, 34 \right. | 012 \left. \right] & 
    %     \operatorname{I} \left[ \, 13 \right. | 024 \left. \right] \\
    %     \operatorname{I} \left[ \, 134 \right. | 02 \left. \right] & 
    %     \operatorname{I} \left[ \, 013 \right. | 24 \left. \right] \\
    % \end{tabular} 
    % \end{align*}
    One of these atoms is shared, and so there are seven atoms
    eliminated in total. 
\item \emph{Minimal optimal prediction}: the forward-time causal states are
	strict functions of the past. They contain no extra information about the
	future that is not contained within the past, but as optimal predictors
	they capture \emph{all} of this information, i.e., all of the excess
	entropy. In information-theoretic terms this means, when conditioning on
	the future, the forward causal states cannot share information with any
	other variables except other forward causal states. The same holds in the
	reverse-time case. For $k, j \in \nats, k \mod 3 = 0$, the disallowed atoms
	are given by:
    \begin{align*}
        \text{(i) For } j \bmod 3 \neq 0 & \text{ and } j > k: \\
        & \Icond{k}{j ; \dots}{k+1, \dots} = 0 \\ 
        \text{(ii) For } j \bmod 3 \neq 1 & \text{ and } j < k+4: \\
        & \Icond{k+4}{j ; \dots}{k+3, \dots} = 0  
    \end{align*}
    This accounts for six variables in each direction.
    % \begin{align*}
    % \def\arraystretch{1.4}
    % \begin{tabular}{D{|}{\mid \,}{-1}@
    %                 {\hskip 2em}
    %                 D{|}{\mid \,}{-1}}
    %     \multicolumn{1}{c}{Forward} & \multicolumn{1}{c}{Reverse} \\
    %     \midrule 
    %     \operatorname{I} \left[ \, 0 2 \right. | 1 3 4 \left. \right] & 
    %     \operatorname{I} \left[ \, 0 4 \right. | 1 2 3 \left. \right] \\
    %     \operatorname{I} \left[ \, 0 4 \right. | 1 2 3 \left. \right] & 
    %     \operatorname{I} \left[ \, 2 4 \right. | 0 1 3 \left. \right] \\
    %     \operatorname{I} \left[ \, 0 2 3 \right. | 1 4 \left. \right] & 
    %     \operatorname{I} \left[ \, 0 1 4 \right. | 2 3 \left. \right] \\
    %     \operatorname{I} \left[ \, 0 2 4 \right. | 1 3 \left. \right] & 
    %     \operatorname{I} \left[ \, 0 2 4 \right. | 1 3 \left. \right] \\
    %     \operatorname{I} \left[ \, 0 3 4 \right. | 1 2 \left. \right] & 
    %     \operatorname{I} \left[ \, 1 2 4 \right. | 0 3 \left. \right] \\
    %     \operatorname{I} \left[ \, 0 2 3 4 \right. | 1 \left. \right] & 
    %     \operatorname{I} \left[ \, 0 1 2 4 \right. | 3 \left. \right] \\
    % \end{tabular} 
    % \end{align*}
    However, two atoms are the same in each direction so there are ten atoms
	eliminated total. 
\item \emph{Markov shielding}: This property does not eliminate any atoms when
    considering only a single time step, but it is worth noting. Since the
	causal states are Markov order-$1$, no information may be shared between
	measurements that is not also contained within the states. For $k \in
	\nats, k \mod 3 = 0$;
    \begin{align*}
        \Icond{k+2}{k+5 ;  \dots }{k+3, k+4, \dots} = 0
		~. 
    \end{align*}
\end{enumerate}

As a final note on indicial ordering, consider the sixth column in
\cref{tab:InfoAtomsModel}, which lists the informational quantities discussed.
Comparing to the fourth and fifth columns, which give the partitioning of
$\RVSet_{\epsilon}$, it is clear that we are able to write the informational
quantities without necessarily including all variables in the conditioning set.
(This is sometimes also true for the joint distribution, but we take it as a
convention to always explicitly include all variables in the joint
distribution.)

We are able to do this because our second property, minimal optimal prediction,
is equivalent to saying that the forward (reverse) causal states render future
(prior) variables conditionally independent with respect to all prior (future)
measurements and prior forward (future reverse) causal states.
\Cref{fig:iDiagramProcessS0S0S1S1} depicts this property as the forward time
causal states covering all space shared between future variables and the prior
measurements and prior forward causal states. 

When writing conditional informational quantities, our convention is to drop all
forward causal states shielded by forward causal states further along in the
future and all reverse causal states shielded by reverse casual states further
in the past. We also drop measurements shielded by causal states in either
direction. To see the result of this, compare the $\overline{\Asubset_i}$ column
in \cref{tab:InfoAtomsModel} to the conditioning variables in the information
quantities listed in the $\atom_i$ column.

\section{Processes}
\label{sec:examples}

With our new information quantities established, we now consider the
exactly-solvable taxonomies for all binary discrete stochastic processes
generated by \eMs with one or two states. These \eMs were enumerated by the
topological \eM enumeration algorithm \cite{John10a}. For one state there is
only an IID process and a constant-value process. For two states, there are
seven unique \eMs, corresponding to four distinct information profiles. The
discrepancy is due to degeneracy in symbol labeling. 

\begin{figure*}
\centering
\includegraphics[width=\textwidth]{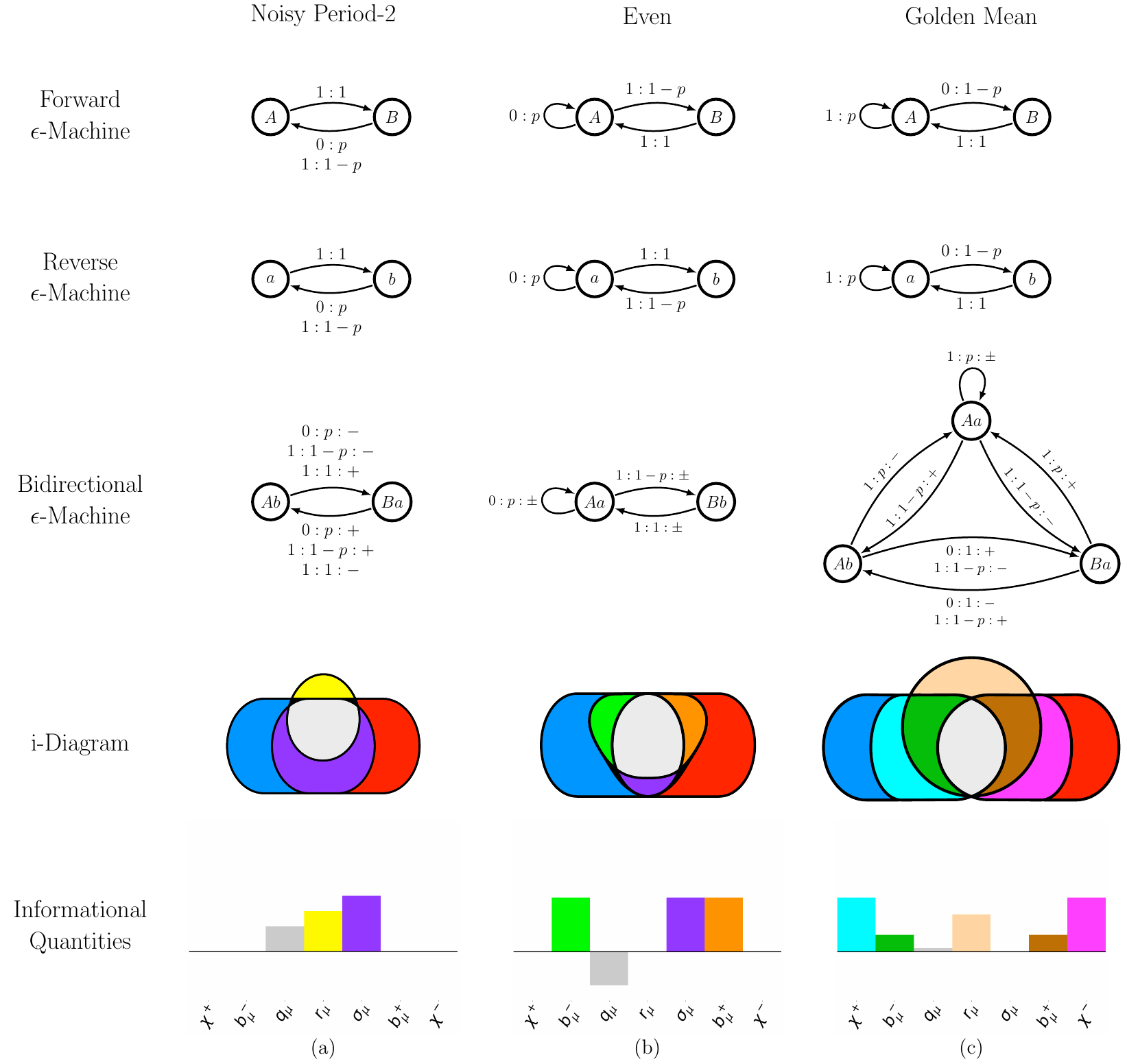}
    \caption{Example prediction taxonomies: The forward \eMs (top row), reverse
	\eMs (second row), bidirectional \eMs (third row), i-diagrams (fourth row),
	and exact informational quantities plotted in a bar chart (bottom row) of
	four discrete stochastic process: a) a Noisy Period$-2$ process, c) the Even
	process, d) the Golden Mean process. Recall that our convention is to use
	uppercase Latin letters for the forward causal states and lower case Latin
	letters for the reverse causal states. }
\label{fig:iDiagramseMachines}
\end{figure*}

\subsection{Independent, Identically-Distributed}
\label{sec:IID}

The first is the simplest possible: an infinite sequence of independent,
identically-distributed (IID) coin flips. The \eM for such a process is given in
\cref{fig:modelMotifs}, as the machine motif for the \typeOne ephemeral
information. In this case, since the process has no structure or memory, there
is only a single causal state in both the forward and reverse directions.

With only a single state the statistical complexity (causal state or model
information) $\Cmu$ vanishes, zeroing out all information in a single
measurement except the \typeOne ephemeral information $\rmu$. If $p=0.5$, $\rmu
= 1$ \unit{\bit}. The information in the infinite past and future diverges. 

\subsection{Periodic}
\label{sec:PeriodN}

The second example process is also a machine motif. An $n$-periodic process
requires exactly $n$ causal states but has only deterministic transitions. As
such, knowledge of the current measurement is equivalent to knowledge of the
infinite past and infinite future, as well as the forward and reverse causal
states. Intuitively, we understand then that the only remaining positive
quantity is $\qmu$. This is the information shared between all model variables.
It is the cycle process' phase information. For an $n$-periodic process, $\qmu
= \log_2 n$ \unit{\bit}. The $n=2$ case is shown in the last row of
\cref{fig:modelMotifs}.

\subsection{Noisy Period-2}
\label{sec:NoisyP2}

The Noisy Period-2 machine is a period-2 process that flips every other bit with
probability $p$. The \eMs are depicted in \cref{fig:iDiagramseMachines} (a). The
forward and reverse machines both have two states. These states can be
identified with each other exactly, $Ab$ and $Ba$, meaning the bidirectional
machine has two states and the processes is noncryptic in both directions.

The Noisy Period-2 process can be seen as a parameterized link between the
Period-2 process at $p=0$, which has one bit of enigmatic information, and a
constant value process when $p=1$, which due to the two (nonminimal) states has
a single bit of elusive information. This allows us to better intuit the
difference between enigmatic and elusive informations---they are both state
informations, differentiated by whether the states are correlated with the
measured bit. It is worth emphasizing that elusive information is not, in
general, only produced by nonminimality; cf. Even process \cref{sec:Even}.
Rather that it can only be isolated in a nonminimal machine. 

For all other values of $p$ the Noisy Period-$2$ has some amount of enigmatic
and elusive information, as well as some \typeOne ephemeral information from
the noisy transition between the two states. 

\subsection{Even}
\label{sec:Even}

The Even process is a binary process of sequences of $0$s of any length
interspersed with even-length sequences of $1$s. Despite the Even process'
simplicity, the process is infinite-order Markov, which is to say that the
probability of the next symbol depends on the infinite length past and cannot be
exactly extrapolated from any finite-length history. As such, there is no finite
Markov model that generates the Even process---it can only be finitely modeled
with a hidden Markov model. 

The probability distribution of the length of the sequences of $0$s and $1$s are
controlled by a single parameter $p \in \left( 0, 1 \right)$.  The \eMs are
depicted in \cref{fig:iDiagramseMachines} (c). There are two forward-time casual
states $\CausalStateSet^+ = \left\{ A, B \right\}$. The self-loop on state $A$
occurs with probability $p$ when the machine is in state $A$. There are two
reverse states, and the forward and reverse states can be identified with each
other exactly, $\left\{ Aa, Bb \right\}$, meaning the Even process is
noncryptic in both directions. There is no ephemeral information due to a lack
of multiple transitions between two states. 

We are left with two \typeOne binding informations, the enigmatic information,
and the elusive information. The entropy rate of the Even process is produced
entirely by the choice between the self-loop and the transition on state $Aa$.
This is, exactly, the \typeOne portion of the forward binding information: it is
not determined by knowledge of the previous state $\ReverseCausalState_0$ (which
due to the noncryptic nature of the process is equivalent to
$\ForwardCausalState_0$). The reverse argument explains the reverse binding
information. 

We can see the presence of the elusive information $\sigmu$ motif in the
transitions between the states in both directions on a $1$. However, the
enigmatic information $\qmu$ is negative, and does not arise from the positive
$\qmu$ motif. How to understand this?

In this case the elusive information is the multivariate mutual information
between $\ReverseCausalState_0 = \ForwardCausalState_0$, $\Present$, and
$\ReverseCausalState_1 = \ForwardCausalState_1$. Recall that the negativity of
multivariate mutual information means that the addition of the third variable
(which can be taken to be any of the three, due to symmetry) \emph{increases}
the shared information between the other two. Notice that the Even state machine
ties one symbol ($0$) to the self loop and one symbol ($1$) to the transition.
This means that knowledge of the measurement reveals that the ordering of the
states is also a structural relationship, increasing the shared information
between the states. 

\subsection{Golden Mean}
\label{sec:GM}

Finally, consider the last informationally distinct two-state binary process.
The Golden Mean Process is a binary process that can have sequences of $1$s of
any length, interspersed with only isolated $0$s. The probability of a $1$s
sequence decreases as the length increases and the nature of this probability
distribution depends on a single parameter $p \in \left(0,1\right)$. The \eMs of
this process family are given in \cref{fig:iDiagramseMachines} (d). There are
two forward-time causal states and $p$ determines the probability split between
the self-loop and the state transition on state $A$, controlling the
probability of seeing a $0$ after a sequence of $1$s. 

The bidirectional machine given in \cref{fig:iDiagramseMachines} shows that the
forward and reverse causal states are not one and the same nor are they
independent---there are three bidirectional causal states. The ``missing''
bidirectional state is $Bb$, which would represent the forward machine being in
state $B$ and the reverse machine being in state $b$ simultaneously. This is
impossible as it implies a sequence of two $0$s. 

Unlike the processes up to this point, the bidirectional machine is
\emph{cryptic}: even if one knows the current causal state in one direction, it
is possible to be uncertain of the current causal state in the opposite
direction. 

The elusive information $\sigmu$ vanishes because the causal state can always be
determined by a measurement of the present ($1$s lead to either $A$ or $a$, $0$s
lead to $B$ or $b$, depending on scan direction). 

All other types of information are represented. The entropy rate splits into
\typeTwo ephemeral information and \typeTwo binding information. We can
intuitively think of this as the new information in each measurement splitting
into a piece that does not explain the future (ephemeral) and a piece that does
(forward binding). Both types of information are of the \typeTwo variety---we
are only uncertain about the observed symbol if we are also uncertain of the
previous reverse causal state $\ReverseCausalState_0$ and the next forward
causal state $\ForwardCausalState_1$. This uncertainty occurs when the machine
is in state $A$, which could transition from $Aa \to Aa$ on a $1$ or from $Ab
\to Ba$ on a $0$. That is, only two of the three possible transitions out of
state $A$, however. The machine can also transition from $Aa \to  Ab$ on a $1$.
This transition is informative about the future, in that it determines the value
of $\ReverseCausalState_1$ and so contributes to the forward binding
information. As usual, this logic also applies in reverse to the reverse binding
information. 

Finally, we have the enigmatic information, which is positive for all values of
$p$. To understand this, we recall our discussion of negative enigmatic
information in the previous example (\cref{sec:Even}). There the value of the
present symbol improved our ability to guess what kind of transition the
machine was undergoing. In this case, the opposite intuition holds. 

\section{Bidirectional Atom Algorithms}
\label{sec:Algorithm}

\EMs are useful not only in that they define a suite of interpretable
informational quantities, but also because knowledge of the \eM allows directly
and exactly calculating those quantities \cite{Crut13a}. With knowledge of a
finitely-specified forward $\eM$ of a discrete stochastic process (which can
even be inferred from time series data \cite{Stre13a}), we can find the reverse
and bidirectional \eMs and from there calculate all the quantities defined in
\cref{sec:anatomyofamodel}. Algorithms to do different aspects of this process
have appeared previously \cite{Crut08b,Jame13a,Crut08a} but we compile those
used here for completeness. 

Before describing the relevant algorithms, we recall and define a few
preliminary concepts. 

A machine $\Machine$ is given by a list of square \emph{transition matrices}
$\left\{ T^{(\meassymbol)} : \meassymbol \in \alphabet \right\}$ where
$T^{(\meassymbol)}_{ij} = \Pr \left( \causalstate_j, \meassymbol \mid
\causalstate_i \right)$. Let $\numFStates = |\ForwardCausalStateSet|$ and
$\numRStates = |\ReverseCausalStateSet|$ so that the transition matrices of the
forward \eM are $\numFStates \times \numFStates$ and the transition matrices of
the reverse \eM are $\numRStates \times \numRStates$.

The \emph{mixed-state algorithm}, fully elucidated in Ref.
% \cite{Jurg20b, Jurg20c},
\cite{Jurg20c},
finds the mixed states $\mixedstate$ of a hidden Markov model $\Machine$.
Briefly, for a length-$\ell$ word $\word$ generated by $\Machine$ the mixed
state $\mixedstate(\word)$ is an observer's best guess as to which state the
machine is in after observing $\word$:
\begin{align}
    \mixedstate(\word) = \left[ \Prob \left( \CausalState_i \mid 
    \MeasSymbol_{0 : \ell} = \word \right) \right]
\end{align}
given an initial guess of $\pi$---the asymptotic stationary distribution of the
machine: $\pi = \pi T$, where the state transition matrix is $T =
\sum_{\meassymbol \in \alphabet} T^{(\meassymbol)}$. The mixed states of a
machine are the set:
\begin{align}
    \MixedStateSet = \left\{ \mixedstate(\word) : \word \in \alphabet^+, 
    \Prob(\word) > 0 \right\}
	~. 
\end{align} 

If the process generated by $\Machine$ has a finite \eM, the mixed-state
algorithm finds the recurrent causal-state set by collecting mixed states
for an arbitrarily long word. In general, $|\MixedStateSet| \to \infty$, so we
typically set a threshold past which if the mixed state set continues to grow,
we assume there is no finite representation. 

\begin{definition}
A \emph{flipped} machine $\flipped{\Machine}$ is a machine where each
transition $T^{(\meassymbol)}_{ij}$ has been replaced with the transition:
\begin{align*}
	\flipped{T}^{(\meassymbol)}_{ji} = T^{(\meassymbol)}_{ij}
	\frac{\pi_{j}}{\pi_i}
	~.
\end{align*} 
This, in effect, flips the direction of the arrows on each transition and
renormalizes the transition probability. This typically produces a nonunifilar
machine.
\end{definition}

\begin{definition}
The \emph{forward switching matrix} $\ForwardS$ between the forward and reverse
\eMs is defined $\ForwardS_{ij} = \Pr( \forwardcausalstate_j |
\reversecausalstate_i)$. The \emph{reverse switching matrix} $\ReverseS$ is
similarly defined $\ReverseS_{ij} = \Pr( \reversecausalstate_i |
\forwardcausalstate_j)$.
\end{definition}

These pieces allow writing down a simple algorithm for \emph{reversing} an
\eM---i.e., constructing the \eM in the reverse direction given the forward
\eM. 

\begin{algorithm}[H]
    \caption{Reverse \eM}
    \begin{algorithmic}[1]
        \Procedure{ReverseEM}{$\ForwardEM$}
        \State \textbf{input} forward \eM $\ForwardEM$.
        \State Flip $\ForwardEM$.
        \State \multiline{Apply the mixed state algorithm to $\flipped{\ForwardEM}$,
        collecting the unique mixed states in a set
        $\MixedStateSet_{\flipped{\ForwardEM}}$. If this set converges to a
        finite set, it consists of the reverse causal states, given in terms of
        a distribution over forward causal states.} 
        \State \multiline{Stack the mixed states vertically into the forward
        switching matrix $\ForwardS$ of shape $\numRStates \times \numFStates$.}
        \State Initialize empty list $T^-$. 
        \For{$\meassymbol$ in $\alphabet$}
            \State Initialize empty $\numRStates \times \numRStates$ matrix 
            ${T^-}^{(\meassymbol)}$. 
            \For{$i = 1, \dots, \numRStates$}
                \State Calculate probability:
                \begin{align*}
                    \mathbf{e}_i \flipped{T^+}^{(\meassymbol)} \mathbf{1} ~.
                \end{align*}
                \State Calculate next state:
                \begin{align*}
                    \frac{\mathbf{e}_i \flipped{T^+}^{(\meassymbol)}}
                    {\mathbf{e}_i \flipped{T^+}^{(\meassymbol)} 
                    \mathbf{1}}  ~.
                \end{align*} 
                \State Initialize empty list.
                \For{$j = 1, \dots, \numFStates$}
                    \If{next state equals $\ForwardS e_j$}
                        \State Append probability to list.
                    \Else
                        \State Append a zero to list.
                    \EndIf
                \EndFor 
            \State Replace the $i$th row of ${T^-}^{(\meassymbol)}$ with list.
            \EndFor
            \State Append ${T^-}^{(\meassymbol)}$ to $T^-$. 
        \EndFor
        \State \multiline{\textbf{return} $\ReverseEM$ as list of reverse 
        \eM transition matrices $T^-$ over symbols $\meassymbol \in \alphabet$.}  
        \EndProcedure
    \end{algorithmic}
\label{alg:reverseEM}
\end{algorithm}

If one starts from the reverse \eM, the forward \eM can be constructed in the
expected manner. Indeed, the labeling of the time direction is somewhat
arbitrary absent a physical system. 

With the forward and reverse \eMs in hand, it is straightforward to construct
the bidirectional machine as in \cref{alg:biEM}. Since retaining consistent
state labeling is important, it is highly recommended to use a data structure
capable of containing labeled axes (rows and columns) and to maintain a
distinct convention for labeling forward and reverse causal states. As already
noted, our convention is to use Latin letters, uppercase for forward states and
lowercase for reverse states. This is particularly important when constructing
the bidirectional machine. 

Let ${\ForwardA}^{(\meassymbol)}$ be the $\numFStates \times \numFStates$
forward-symbol-labeled adjacency matrix of ${T^+}^{(\meassymbol)}$. This is to
say the elements $a^+_ij$ are one when ${T^+}^{(\meassymbol)}_{ij} >0$,
indicating a positive probability of transition, and zero otherwise. 

\begin{algorithm}[H]
\caption{Bidirectional machine}
\begin{algorithmic}[1]
    \Procedure{BidirectionalMachine}{$\ForwardA, \ReverseEM$}
    \State \textbf{input} reverse \eM $\ReverseEM$.
    \State Flip $\ReverseEM$. 
    \State Initialize empty list $T^{\pm}$.
    \For{$\meassymbol$ in $\alphabet$}
        \State From ${\ForwardA}^{(\meassymbol)}$ construct the block matrix:
        \begin{align}
            \begin{bmatrix}
                a^+_{11} \flipped{T^-}^{(\meassymbol)} & \dots
                & a^+_{1\numFStates} \flipped{T^-}^{(\meassymbol)}  \\
                \vdots & & \vdots \\
                a^+_{\numFStates1} \flipped{T^-}^{(\meassymbol)} & \dots 
                & a^+_{\numFStates\numFStates} \flipped{T^-}^{(\meassymbol)} 
            \end{bmatrix} ~,
        \end{align}
        \hspace{2.7em} inheriting state labels as appropriate.
        \State \multiline{Drop all rows and columns consisting of only zeroes, 
        leaving a square matrix.}
        \State \multiline{Append matrix to list of bidirectional machine transition
        matrices $T^{\pm}$.}
    \EndFor 
    \State \multiline{\textbf{return} $\BiEM$ as list of bidirectional machine 
    transition matrices $T^{\pm}$.}  
    \EndProcedure
\end{algorithmic}
\label{alg:biEM}
\end{algorithm}

As with \cref{alg:reverseEM}, the bidirectional machine can be constructed in
the ``reverse direction'', by starting with $\ReverseA$ and $\ForwardEM$ and
making the appropriate substitutions. Regardless, the same bidirectional
machine will be constructed. 

Once the bidirectional machine is in hand, calculating a process' prediction
taxonomy quantities is conceptually straightforward, if somewhat subtle with
regard to tracking indices of the states and observations. See
\cref{alg:infoModel}.

\begin{algorithm}[H]
    \caption{Informational anatomy}
    \begin{algorithmic}[1]
        \Procedure{InfoAnatomyModel}{$\BiEM$} 
    	\State \textbf{input} bidirectional \eM $\BiEM$.
        \State \multiline{Generate list of nonzero measure partitions, 
        according to the indicial rules laid out in 
        \cref{sec:StructureineMInfoAtoms}.}
        \State \multiline{Calculate the probability of all possible 
        transitions of the bidirectional machine from an initial distribution
		over states. Unless otherwise noted, use the stationary distribution 
        $\pi^{{\pm}}$.}
        \State Initialize empty list. 
        \For{$\Asubset_i$ in partition}
            \State Apply the information function \cref{eq:generalInfoFunction}.
            \State Append information value to list.
        \EndFor
        \State \textbf{return} list of information quantities.
        \EndProcedure
    \end{algorithmic}
    \label{alg:infoModel}
\end{algorithm}

Once again, data structures capable of retaining labeled axes are recommended,
along with a consistent indicial labeling strategy as laid out in
\cref{sec:StructureineMInfoAtoms}.

\section{Conclusion}
\label{sec:conclusion}

This concludes our development of the informational taxonomy of an optimally
predicted and retrodicted process. There are several few points of interest
to highlight. 

Step 3 of \cref{alg:infoModel} requires choosing a distribution over the states
of the bidirectional machine to determine the probability of paths through the
machine (and, of observing words of the process). We have not discussed this
aspect of the prediction taxonomy explicitly, implicitly assuming that the
process is in the stationary distribution. However, this is a choice, and a
potentially interesting one---one can calculate the taxonomy of information
measures for any distribution over the states of the bidirectional machine.
That said, the canonical computational mechanics quantities like $\Cmu$ are typically defined in terms of the stationary distribution $\pi$
% \cite{Crut01a,Jame11a}. 
\cite{Jame11a}. 

As the \eMs are constrained to be ergodic Markov chains over the states, any
initial distribution will eventual converge to the stationary distribution when
evolved by the state transition matrix $T$. We conjecture this is true for the
bidirectional machine as well, so one can track the convergence of the
prediction taxonomy quantities by starting the bidirectional machine away
from equilibrium and allowing it to evolve towards the stationary distribution. 

Another, alternative analysis is to explore the informational
properties of prediction when the machine is constrained to a subset of possible
observations. The informational exploration of the \eM operating away from the
stationary state is an intriguing area of exploration that has been considered
in related work on thermodynamically coupled \eMs \cite{Crut16a}. We reserve the
discussion of this avenue for future work. 

We also wish to note that this development is closely related to other
fine-grained informational analyses of stochastic processes. In particular, we
are interested in exploring the relationship between the results here and
from the partial information decomposition \cite{Beer14a}. Reference
\cite{Jame11a} showed that analyzing the quantities described in
\cref{sec:ProcessInfoAtoms} with the partial information lattice allows one to
relate enigmatic information $\qmu$ to the synergy and redundancy. We are
interested in a similar analysis with our new, expanded taxonomy, but this is
outside the present scope.

It is also important to note our focus on irreducible information measures is
not intended to exclude the use of aggregate information measures or disregard
their importance in informational analysis of processes. Rather our goal was an
algorithmically calculable suite of measures that is consistent across
processes and span the space of Shannon measures. However, there are many
multivariate measures of interest---total correlation \cite{Watanabe60}, dual
total correlation \cite{Han75a}, G\'as-K\"orner common information
\cite{Gacs73a}, among many others. Reference \cite{Jame16a} contains a helpful
list for three variables. Their relationship to the more basic Shannon measures
is of interest.

As one may conclude from the indicial rules laid out in
\cref{sec:StructureineMInfoAtoms} and \cref{alg:infoModel}, the procedure for
generating the informational anatomy of a model can be straightforwardly
extended beyond assuming the present is single time step rather than, say, a
block of finite duration. Indeed, doing so leads to even more intriguing
informational representations of processes and complexity measures.  However,
this extension too is beyond the present scope, but will be discussed instead
in a sequel.

\section*{Acknowledgments}

The authors thank Ryan James for insights and helpful comments. The authors
also thank the Telluride Science and Innovation Center for hospitality during
visits and the participants of the Information Engines Workshops there. This
material is based upon work supported by, or in part by, U.S. Army Research
Laboratory and the U.S. Army Research Office under grant W911NF-21-1-0048.

\bibliographystyle{unsrtnat}
\bibliography{jurgens_master}
% \bibliography{chaos}

\end{document}